\documentclass[12pt]{article}
\usepackage{amsmath,amssymb,theorem,cite,epsfig,url,psfrag,eepic,mathtools,amsmath,mathtools,graphicx,xcolor}
\usepackage{hyperref}
\usepackage{graphicx}
\usepackage{epigraph}
\advance \topmargin by -\headheight
\advance \topmargin by -\headsep     
\evensidemargin \oddsidemargin
\def\Maketitle{{\def\newpage{}\maketitle}}
\makeatletter
\def\Appendix{\appendix
	\def\@seccntformat##1{Appendix~\csname the##1\endcsname.~~}}
\makeatother
\makeatletter
\@addtoreset{equation}{section}

\makeatother
\makeatletter
\newcommand{\zerounderset}[3][\mathord]{%
	#1{\vtop{
			\let\\\cr
			\baselineskip\z@skip\lineskip.25ex
			\ialign{\hidewidth$##$\hidewidth\crcr
				\omit$#3$\cr
				#2\crcr
			}%
	}}%
}
\makeatother
\begin{document}
\title{\textbf{Integrable structure of \textrm{BCD} conformal field theory \\ and boundary Bethe ansatz for affine Yangian}\vspace*{.3cm}}
\date{}
\author{Alexey Litvinov$^{1,2}$ and Ilya Vilkoviskiy$^{2,3}$\\[\medskipamount]
\parbox[t]{0.85\textwidth}{\normalsize\it\centerline{1. Landau Institute for Theoretical Physics, 142432 Chernogolovka, Russia}}\\
\parbox[t]{0.85\textwidth}{\normalsize\it\centerline{2. Center for Advanced Studies, Skolkovo Institute of Science and Technology, 143026 Moscow, Russia}}\\
\parbox[t]{0.85\textwidth}{\normalsize\it\centerline{3. National Research University Higher School of Economics, 119048 Moscow, Russia}}}
\Maketitle
\begin{abstract}
In these notes we study integrable structures of conformal field theory with \textrm{BCD} symmetry. We realise these integrable structures as affine Yangian $\mathfrak{gl}(1)$ "spin chains" with boundaries. We provide three solutions of Sklyanin \textrm{KRKR} equation compatible with affine Yangian $R$-matrix and derive Bethe ansatz equations for the spectrum.
\end{abstract}
\section{Introduction} 

The study of integrable structure of conformal field theory began with the seminal series of papers of Bazhanov, Lukyanov and Zamolodchikov \cite{Bazhanov:1994ft,Bazhanov:1996dr,Bazhanov:1998dq} devoted to study of quantum KdV integrable system. In particular, the set of generating functions for local and non-local Integrals of Motion has been explicitly constructed. Unfortunately the construction of \cite{Bazhanov:1994ft,Bazhanov:1996dr,Bazhanov:1998dq} does not known to provide by itself any equations for the spectrum  of the Integrals of Motion. However, using the similarity with the theory of ordinary differential equations and bunch of analytic intuition, the same authors in \cite{Bazhanov:2004fk} were able to express the spectrum of the local IM's  in terms of the solutions of certain algebraic system of equations. Later these equations were generalized for some other integrable structures, such as Fateev models or quantum AKNS model (see \cite{Kotousov:2019nvt} for the list of all known cases). Despite the obvious success of BLZ program, it is still unclear where the algebraic equations of \cite{Bazhanov:2004fk} come from, and whether they can be easily generalized for other models of CFT.

Recently it becomes clear that there is parallel approach based on the affine Yangian symmetry.  The advantage of this approach is that it fits in general framework of the quantum inverse scattering method and provides Bethe ansatz equations for the spectrum.  Being originally formulated geometrically \cite{varagnolo2000quiver,Nakajima:fk,Maulik:2012wi}, it can be rephrased entirely algebraically in CFT terms\footnote{For the modern review of the geometric approach and more advanced topics see Andrei Okounkov's summer lecture course \href{https://sites.google.com/view/andrei-okounkov-lecture-course/home}{\texttt{sites.google.com/view/andrei-okounkov-lecture-course/home}}.}.  In  \cite{Litvinov:2020zeq}, using this algebraic approach, we studied the integrable structures in CFT related to $\textrm{Y}\big(\widehat{\mathfrak{gl}}(1)\big)$, the affine Yangian of $\mathfrak{gl}(1)$ \cite{Tsymbaliuk:2014fvq}. These integrable structures appear in $W-$algebras of $\mathrm{A}_n$ type and its super-algebra  generalizations and can be viewed as twist deformations of the quantum Gelfand-Dikii hierarchies (quantum ILW type integrable systems).  We used $\textrm{RLL}$ formulation of $\textrm{Y}\big(\widehat{\mathfrak{gl}}(1)\big)$, where $\textrm{R}$ stands for the Maulik-Okounkov $R-$matrix \cite{Maulik:2012wi}, explicitly constructed off-shell Bethe vectors and have shown that on Bethe ansatz equations these vectors diagonalize both \textrm{KZ} and local Integrals of Motion.

In current notes we generalize the results and the methods of  \cite{Litvinov:2020zeq} to the $W$-algebras of $\textrm{BCD}$ type. The key new ingredient, which appears in this case is the analog of Sklyanin's $K-$matrix \cite{Sklyanin:1988yz}, introduced by him for studying of spin chains with boundary. The "boundary" in the current context corresponds to the endpoints of the affine Dynkin diagram for a given integrable system. This fact has been already noticed and studied in trigonometric case in \cite{feigin2020deformations}. Here we restrict ourselves to the conformal case, but consider  the problem of diagonalization of Integrals of Motion. Similar to the $\textrm{A}$ case \cite{Litvinov:2020zeq}, it is convenient to diagonalize \textrm{KZ} Integrals of Motion (called reflection operators in \cite{Kotousov:2019nvt}) rather that local ones. We explicitly construct the off-shell Bethe vector, which depends on auxiliary parameters $x_1,\dots,x_N$, where $N$ is the level, and show that the \textrm{KZ} operator acts diagonally on this vector provided that $x_k$'s satisfy Bethe ansatz equations. These equations (formula \eqref{BAE}) together with the explicit form of off-shell Bethe vector (formula \eqref{definition-off-shell-bethe-vector}) constitute the main results of our paper.

This paper is organized as follows. In section \ref{integrable-systems} we introduce integrable systems of $\mathrm{BCD}$ type, as a commutant of affine system of screening operators. We also provide an explicit form of first non-trivial local Integral of Motion. In section \ref{RK-matrices} we review the Maulik-Okounkov $R-$matrix, introduce the notion of the Sklyanin $K-$matrix in this context and review basic facts about affine Yangian of $\widehat{\mathfrak{gl}}(1)$. Section \ref{off-shell-introduction} is devoted to explicit construction of off-shell Bethe vectors. In section \ref{BA-proof} we explicitly diagonalize \textrm{KZ} Integral of Motion. In section \ref{concl} we give some concluding remarks, in particular we give the conjecture for eigenvalues of local Integrals of Motion. We also provide some details on integrable systems of low rank, in particular for Bullough-Dodd model. In appendices we provide supplementing calculations and useful formulae.
\section{Integrable systems of \texorpdfstring{$\mathrm{BCD}$}{BCD} type in CFT}\label{integrable-systems} 
The integrable systems studied in this paper can be realized by the $n-$component bosonic free field $\boldsymbol{\varphi}=(\varphi_1,\dots,\varphi_n)$. Local Integrals of Motion have the following general form
\begin{equation}\label{local-IMs}
  \mathbf{I}_s=\frac{1}{2\pi}\int_{0}^{2\pi}G_{s+1}(z)dz,\qquad
  \bar{\mathbf{I}}_s=\frac{1}{2\pi}\int_{0}^{2\pi}\bar{G}_{s+1}(\bar{z})d\bar{z},
\end{equation}
where $G_{s+1}(z)$  and $\bar{G}_{s+1}(\bar{z})$ are the local densities with the spins $s$ belonging to some set, which is a characteristic property of a particular integrable system. The important property of local IM's is that they form the commutative set
\begin{equation}\label{II-comm}
    [\mathbf{I}_r,\mathbf{I}_s]=0.
\end{equation}

The best way to describe our integrable systems goes through affine Toda QFT
\begin{equation}\label{Affine-Toda-action}
S=\int\Big(\frac{1}{8\pi}(\partial_a\boldsymbol{\varphi}\cdot\partial_a\boldsymbol{\varphi})+\Lambda\sum_{r=0}^{n}e^{b(\boldsymbol{\alpha}_r\cdot\boldsymbol{\varphi})}\Big)d^2z.
\end{equation}
where the vectors $(\boldsymbol{\alpha}_0,\boldsymbol{\alpha}_1,\dots,\boldsymbol{\alpha}_n)$ have the Gram matrix corresponding to the one of the affine Dynkin diagrams of $\mathrm{BCD}$ type:
\begin{center}
\psfrag{d}{$\widehat{\mathrm{D}}_n$}
\psfrag{b}{$\widehat{\mathrm{B}}_n$}
\psfrag{bb}{$\widehat{\mathrm{B}}_n^{\vee}$}
\psfrag{c}{$\widehat{\mathrm{C}}_n$}
\psfrag{cc}{$\widehat{\mathrm{C}}_n^{\vee}$}
\psfrag{bc}{$\widehat{\mathrm{BC}}_n$}
\includegraphics[width=.8\textwidth]{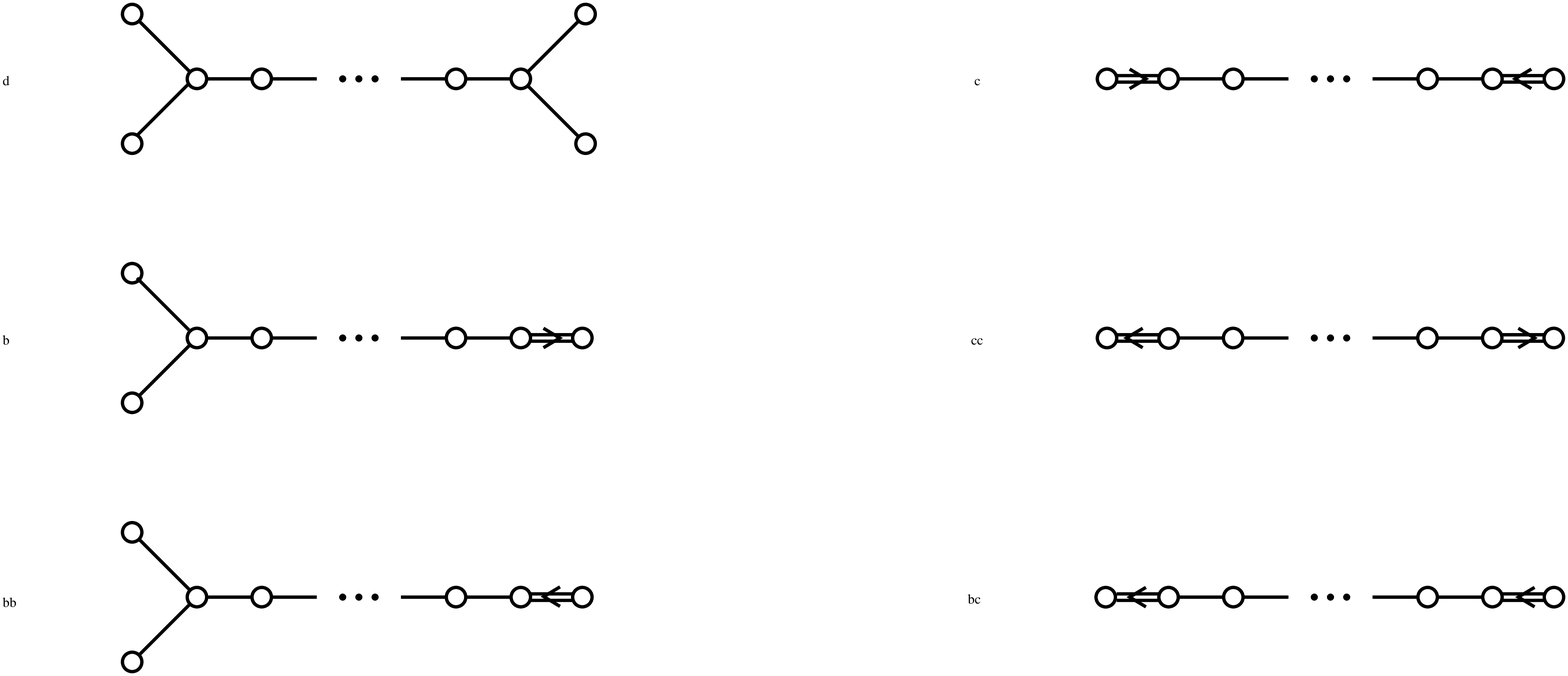}
\end{center}
and $b$ is the coupling constant. Using the standard parametrization for the roots one can express the scalar products in the exponents in \eqref{Affine-Toda-action} as
\begin{equation}\label{roots}
    (\boldsymbol{\alpha}_0\cdot\boldsymbol{\varphi})=
      \begin{cases}
        -\varphi_1\\
        -2\varphi_1\\
        -\varphi_1-\varphi_2
      \end{cases}\quad
  (\boldsymbol{\alpha}_r\cdot\boldsymbol{\varphi})=\varphi_r-\varphi_{r+1}\quad\text{for}\quad 0<r<n,\quad
    (\boldsymbol{\alpha}_n\cdot\boldsymbol{\varphi})=
    \begin{cases}
    \varphi_n\\
    2\varphi_n\\
        \varphi_{n-1}+\varphi_n
    \end{cases}
\end{equation}
That is each of the affine diagrams can be interpreted as non-affine $\mathrm{A}_{n-1}$ diagram with two boundary conditions which can be of three types $\mathrm{B}$, $\mathrm{C}$ or $\mathrm{D}$ corresponding to the short root, the long root or the root of the length $\sqrt{2}$ correspondingly.  

The theories \eqref{Affine-Toda-action} are known to be integrable both classically and quantum mechanically.  They share an interesting property of the duality (see e.g. \cite{Corrigan:1994nd}). Namely, both $\widehat{\mathrm{D}}_n$ and $\widehat{\mathrm{BC}}_n$ theories are self-dual with respect to the substitution $b\rightarrow b^{-1}$, while $\widehat{\mathrm{B}}_n$ and $\widehat{\mathrm{B}}^{\vee}_n$ as well as $\widehat{\mathrm{C}}_n$ and $\widehat{\mathrm{C}}^{\vee}_n$ are mapped to each other. The quantum integrability implies that the theory admits the set of local Integrals of Motion whose short distance limit coincides with $\mathbf{I}_s$ and $\bar{\mathbf{I}}_s$ from \eqref{local-IMs}.  

The integrals  $\mathbf{I}_s$ and $\bar{\mathbf{I}}_s$ by themselves can be defined up to a total factor from the equation (and similar antiholomorphic equation)
\begin{equation}\label{commutant-with-screening}
   \frac{1}{2\pi i}\oint_{\mathcal{C}_z}e^{b\left(\boldsymbol{\alpha}_r\cdot\boldsymbol{\varphi}(\xi)\right)}G_{s+1}(z)d\xi=\partial V_{s}(z),
\end{equation}
where $V_s(z)$ is some local field (and similar formula for $\bar{G}_{s+1}$). Using \eqref{commutant-with-screening} one can construct first few local IM's explicitly. It is convenient to write them in Nekrasov epsilon notations\footnote{The answer will depends only on the ratio $\frac{\epsilon_1}{\epsilon_2}$, so without loss of generality we may assume $\epsilon_1 \epsilon_2=1$ and thus
\begin{equation}\label{epsilon-notations-2}
  b=\epsilon_2,\qquad b^{-1}=\epsilon_1\quad\text{and}\quad
  Q=b+\frac{1}{b}=-\epsilon_3.
\end{equation}}
\begin{equation}\label{epsilon-notations}
  b=\sqrt{\frac{\epsilon_2}{\epsilon_1}},\qquad b^{-1}=\sqrt{\frac{\epsilon_1}{\epsilon_2}}\quad\text{and}\quad
  \epsilon_3\overset{\text{def}}{=}-\epsilon_1-\epsilon_2.
\end{equation} 
The first non-trivial local Integral of Motion is $\mathbf{I}_3$ and the corresponding Wick ordered density has the form
\begin{multline}\label{G4-density}
   G_4(z)=\big(\partial\boldsymbol{\varphi}\cdot\partial\boldsymbol{\varphi}\big)^2-\frac{1}{3}\left(2n-\frac{\epsilon_{\alpha}+\epsilon_{\beta}}{\epsilon_3}\right)\sum_{k=1}^n\big(\partial\varphi_k\big)^4+\\+
   \frac{4\epsilon_3}{\sqrt{\epsilon_1\epsilon_2}}\sum_{k=1}^n\partial\varphi_k^2\left(\sum_{j<k}\Big(j-1+\frac{\epsilon_3-\epsilon_{\alpha}}{2\epsilon_3}\Big)\partial^2\varphi_j-\sum_{j>k}\Big(n-j+\frac{\epsilon_3-\epsilon_{\beta}}{2\epsilon_3}\Big)\partial^2\varphi_j\right)+\\+\left(2n+\frac{4(n-1)(\epsilon_1^2+\epsilon_2^2)}{3\epsilon_1\epsilon_2}+\frac{(\epsilon_1\epsilon_2-2\epsilon_3^2)(\epsilon_{\alpha}+\epsilon_{\beta}-2\epsilon_3)}{3\epsilon_1\epsilon_2\epsilon_3}\right)\big(\partial^2\boldsymbol{\varphi}\cdot\partial^2\boldsymbol{\varphi}\big)-\\-
   \frac{4\epsilon_3^2}{\epsilon_1\epsilon_2}\sum_{i\leq j}
   \left(i-1+\frac{\epsilon_3-\epsilon_{\alpha}}{2\epsilon_3}\right)
   \left(n-j+\frac{\epsilon_3-\epsilon_{\beta}}{2\epsilon_3}\right)(2-\delta_{ij})\partial^2\varphi_i\partial^2\varphi_j,
\end{multline}
where each of the indexes $\alpha$ and $\beta$ takes the values $1$, $2$ and $3$, corresponding to either $\mathrm{B}$, $\mathrm{C}$ or $\mathrm{D}$ boundary conditions. 

We stress that in general the solution to the commutativity equation \eqref{commutant-with-screening} should be searched in terms of analytically regularized  densities rather that Wick ordered ones. In the case of the density of spin $4$ these two differ by an amount which is by itself an Integral of Motion. In general this is not the case and starting from the spin $6$ one expects to have corrections to the Wick ordered density, which formally correspond to lower spins (see section \ref{concl} for the example).
\section{Maulik-Okounkov \texorpdfstring{$R$}{R}-matrix, \texorpdfstring{$K$}{K}-matrix}\label{RK-matrices}
The Maulik-Okounkov $R-$matrix is related to the Liouville reflection operator \cite{Zamolodchikov:1995aa} as \begin{equation}\label{Rij}
  \mathcal{R}_{ij}=\mathcal{R}[\partial\varphi_i-\partial\varphi_j].
\end{equation}
We will use both notations $\eqref{Rij}$ interchangeably.  Sometimes it may also be convenient to use the notation $\mathcal{R}_{i,j}(u_i-u_j)$ in order to emphasise the value of the zero mode (see \eqref{free-field}).

This reflection operator can be defined up to a normalisation factor from the condition ($Q=\frac{\epsilon_1+\epsilon_2}{\sqrt{\epsilon_1\epsilon_2}}$)
\begin{equation}\label{RLL-Miura}
  \mathcal{R}[\partial\varphi_i-\partial\varphi_j](Q\partial-\partial\varphi_i)(Q\partial-\partial\varphi_j)=(Q\partial-\partial\varphi_j)(Q\partial-\partial\varphi_i)\mathcal{R}[\partial\varphi_i-\partial\varphi_j],
\end{equation}
where $\varphi_k$ is the free bosonic field
\begin{equation}\label{free-field}
\partial\varphi_k(x)=-i\frac{u_k}{\sqrt{\epsilon_1\epsilon_2}}+\sum\limits_{n\ne 0}a^{(k)}_n e^{-i n x},\qquad
 [a^i_{m},a^j_{n}]=m\delta_{m,-n}\delta_{i,j}.
\end{equation}

In order to introduce the $K$-operator, we consider rank two $W$ algebras of $\textrm{BCD}$ type. They can be defined as commutants of screening operators (here $b=\frac{\epsilon_2}{\sqrt{\epsilon_1\epsilon_2}}$)
\begin{equation}\label{Screening-rank-2}
    \mathcal{S}_1=\int e^{b(\varphi_1-\varphi_2)}dz,\qquad \mathcal{S}_2=
    \begin{cases}
    \int e^{b\varphi_2}dz\;\;\text{for}\;\mathrm{B},\\
    \int e^{2b\varphi_2}dz\;\;\text{for}\;\mathrm{C},\\
    \int e^{b(\varphi_1+\varphi_2)}dz\;\;\text{for}\;\mathrm{D}.
    \end{cases}
\end{equation}
The corresponding holomorphic currents $W_2$ and $W_4$ have the explicit  form
\begin{equation}\label{W2}
  W_2=(\partial\varphi_1)^2+(\partial\varphi_2)^2+\frac{2\epsilon_3}{\sqrt{\epsilon_1\epsilon_2}}\partial^2\varphi_1+\frac{\epsilon_3-\epsilon_{\alpha}}{\sqrt{\epsilon_1\epsilon_2}}(\partial^2\varphi_2+\partial^2\varphi_1)
\end{equation}
and
\begin{multline}\label{W4}
   W_4=(\partial\varphi_1)^2(\partial\varphi_2)^2+\frac{2\epsilon_3}{\sqrt{\epsilon_1\epsilon_2}}\partial\varphi_1\partial\varphi_2\partial^2\varphi_2+\frac{\epsilon_3-\epsilon_{\alpha}}{\sqrt{\epsilon_1\epsilon_2}}\big((\partial\varphi_1)^2\partial^2\varphi_2+(\partial\varphi_2)^2\partial^2\varphi_1\big)-\\-
   \frac{\epsilon_{3}\epsilon_{\alpha}}{\epsilon_1\epsilon_2}(\partial^2\varphi_1)^2+\frac{(\epsilon_3-\epsilon_{\alpha})^2}{\epsilon_1\epsilon_2}\partial^2\varphi_1\partial^2\varphi_2-\frac{(\epsilon_1-\epsilon_{\alpha})(\epsilon_2-\epsilon_{\alpha})}{2\epsilon_1\epsilon_2}\left(\partial\varphi_1\partial^3\varphi_1+\partial\varphi_2\partial^3\varphi_2\right)-\\
   -\frac{\epsilon_3(\epsilon_3-\epsilon_{\alpha})}{\epsilon_1\epsilon_2}\left(\partial\varphi_1\partial^3\varphi_1-\partial\varphi_1\partial^3\varphi_2\right)+
   \frac{\epsilon_3}{\sqrt{\epsilon_1\epsilon_2}}\left(\frac{\epsilon_{\alpha}(\epsilon_3-\epsilon_{\alpha})}{2\epsilon_1\epsilon_2}-\frac{\epsilon_3^2}{\epsilon_1\epsilon_2}-\frac{1}{3}\right)\partial^4\varphi_1
\end{multline}
where $\alpha=1,2,3$ correspond to $\mathrm{B}$, $\mathrm{C}$ and $\mathrm{D}$ $W$ algebras correspondingly.

Each screening operator \eqref{Screening-rank-2} generates the reflection operator according to the rule
\begin{equation}\label{definition-KR}
    \mathcal{R}_{1,2} W_s=W_s\Bigg|_{\varphi_{1} \leftrightarrow \varphi_{2}}\hspace*{-25pt}\mathcal{R}_{1,2},\qquad
   \mathcal{K}_2 W_s=W_s\Bigg|_{\varphi_2 \to -\varphi_2}\hspace*{-25pt} \mathcal{K}_2,
\end{equation}
for $s=2,4$. We have $\mathcal{R}_{1,2}=\mathcal{R}[\partial\varphi_1-\partial\varphi_2]$, while $\mathcal{K}_2$ is also equal to the reflection operator of the re-scaled argument
\begin{align} 
\label{K1}
\mathcal{K}_2^1=\mathcal{R}[\sqrt{2}\partial\varphi_2]\Big|_{\epsilon_1\to \sqrt{2}\epsilon_1,\epsilon_2\to \epsilon_2/\sqrt{2}} \,\quad &\text{for $\mathrm{B}$ series}\\ \label{K2}
\mathcal{K}_2^2=\mathcal{R}[\sqrt{2}\partial\varphi_2]\Big|_{\epsilon_1\to \epsilon_1/\sqrt{2},\epsilon_2\to \sqrt{2}\epsilon_2} \,\quad &\text{for $\mathrm{C}$ series}\\  \label{K3}
\mathcal{K}_2^3=\textrm{Id} \,\quad &\text{for $\mathrm{D}$ series}
\end{align}
Note that $\mathcal{K}_2^3=\textrm{Id}$ is the simplest among the operators, as it does not depend on spectral parameter and has very simple action on bosons.

Now, similar to the argument of Maulik and Okounkov, the $K$-operator obeys Sklyanin's $\textrm{KRKR}$ equation\footnote{Let us note that there is more convenient form of $\textrm{KRKR}$ equation used by Sklyanin \cite{Sklyanin:1988yz}:
\begin{equation}\label{KRKR-equation} 
    \mathcal{R}_{{1},{2}}(u_1-u_2)\tilde{\mathcal{K}}_1(u_1)\mathcal{R}_{{2},1}(u_2+u_1)\tilde{\mathcal{K}}_2(u_2)=\tilde{\mathcal{K}}_2(u_2)\mathcal{R}_{{1},2}(u_1+u_2)\tilde{\mathcal{K}}_1(u_1)\mathcal{R}_{2,1}(u_1-u_2).
\end{equation}
These two equations actually differ by the redefinition of $K-$operator and overall conjugation by the reflection of bosonic modes $a^{1,2}_n\to -a^{1,2}_n$ , $n\ne 0$}
\begin{equation}\label{reflection-equation}
\mathcal{R}[\partial\varphi_1-\partial\varphi_2]\mathcal{K}^{\alpha}_1\mathcal{R}[\partial\varphi_1+\partial\varphi_2]\mathcal{K}^{\alpha}_2=\mathcal{K}^{\alpha}_2\mathcal{R}[\partial\varphi_1+\partial\varphi_2]\mathcal{K}^{\alpha}_1\mathcal{R}[\partial\varphi_1-\partial\varphi_2].
\end{equation}
It is interesting to note that $\mathcal{K}^1$, $\mathcal{K}^2$ and $\mathcal{K}^3$ seem to exhaust all solutions to $\textrm{KRKR}$ equation \eqref{reflection-equation} which preserve the grading operator $\int W_2dz$. This is an unproven statement, confirmed by explicit calculations on lower levels.
\subsection{KZ integrals of motion.}
Having defined $R-$ and $K-$operators, one can define the important family of IOM's constructed from two solutions of $\textrm{KRKR}$ equation -- the so called \textrm{KZ} Integrals of Motion. Let us introduce the following operators:
\begin{gather} 
\mathcal{T}^+_i=\mathcal{R}_{i,\overline{i+1}}\dots\mathcal{R}_{i,\overline{n}}\mathcal{K}^{\alpha}_i \mathcal{R}_{i,n} \dots \mathcal{R}_{i,i+1},\\
\mathcal{T}^-_i=\mathcal{R}_{i,1}\dots\mathcal{R}_{i,i-1}\mathcal{K}^{\beta}_{i}\mathcal{R}_{1,\bar{i}}\dots \mathcal{R}_{i-1,\bar{i}},\\ \label{KZ_oper}
\mathcal{I}^{\textrm{KZ}}_i=\mathcal{T}^-_i\mathcal{T}^+_i
\end{gather}
where we defined the conjugation operator $D_i$
\begin{gather}
D_i f(\boldsymbol{\varphi})=f(\boldsymbol{\varphi})\Big|_{\varphi_i\to -\varphi_i} D_i,\\
\mathcal{R}_{i,\bar{j}}=D_j\mathcal{R}_{i,j}D_j=\mathcal{R}[\partial\varphi_i+\partial\varphi_j],\\
\mathcal{R}_{\bar{i},j}=D_i\mathcal{R}_{i,j}D_i=\mathcal{R}[-\partial\varphi_i-\partial\varphi_j],
\end{gather}
Using $\textrm{KRKR}$ equation \eqref{reflection-equation}, it is straightforward to check that 
\begin{equation}
    [\mathcal{I}^{\textrm{KZ}}_i,\mathcal{I}^{\textrm{KZ}}_j]=0.
\end{equation}

It is also possible to prove the commutativity of \textrm{KZ} Integrals of Motion and local ones. Indeed, any screening operator $S_{\alpha}$ acts non-trivially only in two (or one at the endpoints) spaces. In order to point it out we will equip it with the label $i$, such that $S_{\alpha_i}$ acts in the space of two bosons $\mathcal{F}_i\otimes \mathcal{F}_{i+1}$ for $i\ne 0,n$, while $S_{\alpha_0}$ and $S_{\alpha_n}$ acts only on first and the last boson correspondingly (see \eqref{roots}). Now, from the very definition of the reflection operators \eqref{definition-KR} any operator $\mathcal{O}_i$ which commutes with $S_{\alpha_i}$ has a nice intertwining property with the reflection operators
\begin{align}
    \mathcal{R}_{i,i+1} \mathcal{O}_i&=\mathcal{O}_i\Big|_{\varphi_{i} \leftrightarrow \varphi_{i+1}}\mathcal{R}_{i,i+1} \,,\quad i=1\dots n-1\\
  \mathcal{K}_i \mathcal{O}_i&=\mathcal{O}_i\Big|_{\varphi_i \to -\varphi_i}\ \ \ \ \mathcal{K}_i \,,\quad i=0,n.
\end{align}
As local IM's commute with all screening operators, they nicely intertwine with both $\mathcal{T}^-$ and $\mathcal{T}^+$
\begin{equation}
\mathcal{T}_i^+\mathbf{I}_s=\mathbf{I}_s\Big|_{\varphi_i\to -\varphi_i}\mathcal{T}_i^+,\qquad
\mathcal{T}_i^- \ \mathbf{I}_s\Big|_{\varphi_i\to -\varphi_i}=\mathbf{I}_s\ \mathcal{T}_i^-,
\end{equation}
which proves the commutativity  $[\mathbf{I}_s,\mathcal{I}_i^{\textrm{KZ}}]=0$.
\subsection{Review of the Affine Yangian \texorpdfstring{$\mathrm{Y}(\widehat{\mathfrak{gl}}_1)$}{Y(gl(1))}}
Let us remind the basic properties of $\textrm{RLL}$ algebra and it's equivalent description in terms of generating currents $h$, $e$ and $f$ (for more details see \cite{Litvinov:2020zeq}).

The Maulik-Okounkov $R$-matrix defines the Yang-Baxter algebra ($\textrm{YB}\bigl(\widehat{\mathfrak{gl}}(1)\bigr)$) in the standard way
\begin{equation}\label{YB-algebra}
 \mathcal{R}_{ij}(u-v)\mathcal{L}_{i}(u)\mathcal{L}_{j}(v)=\mathcal{L}_{j}(v)\mathcal{L}_{i}(u)\mathcal{R}_{ij}(u-v).
\end{equation}
Here $\mathcal{L}_i(u)$ is treated as an operator in some quantum space, a tensor product of $n$ Fock spaces in our case, and as a matrix in the auxiliary Fock space $\mathcal{F}_u$. The algebra \eqref{YB-algebra} becomes an infinite set of quadratic relations between the matrix elements labeled by two partitions
\begin{equation}
 \mathcal{L}_{\scriptscriptstyle{\boldsymbol{\lambda},\boldsymbol{\mu}}}(u)\overset{\text{def}}{=}\langle u|a_{\boldsymbol{\lambda}}\mathcal{L}(u)a_{-\boldsymbol{\mu}}|u\rangle\quad\text{where}\quad
 a_{-\boldsymbol{\mu}}|u\rangle=a_{-\mu_1}a_{-\mu_2}\dots|u\rangle.
\end{equation}

Let us introduce three basic currents of degree $0$, $1$ and $-1$
\begin{equation}\label{efh-def}
 h(u)\overset{\text{def}}{=}\mathcal{L}_{\scriptscriptstyle{\varnothing,\varnothing}}(u),\qquad
 e(u)\overset{\text{def}}{=}h^{-1}(u)\cdot \mathcal{L}_{\scriptscriptstyle{\varnothing,\Box}}(u)\quad\text{and}\quad
 f(u)\overset{\text{def}}{=}\mathcal{L}_{\scriptscriptstyle{\Box,\varnothing}}(u)\cdot h^{-1}(u),
\end{equation}
as well as an auxiliary current (as we will see \eqref{hh-relation} it  also belongs to the Cartan subalgebra of $\textrm{YB}\bigl(\widehat{\mathfrak{gl}}(1)\bigr)$)
\begin{equation}\label{psi-definition}
 \psi(u)\overset{\text{def}}{=}\Bigl(\mathcal{L}_{\scriptscriptstyle{\Box,\Box}}(u+\epsilon_3)-\mathcal{L}_{\scriptscriptstyle{\varnothing,\Box}}(u+\epsilon_3)h^{-1}(u+\epsilon_3)
 \mathcal{L}_{\scriptscriptstyle{\Box,\varnothing}}(u+\epsilon_3)\Bigr)h^{-1}(u+\epsilon_3)
\end{equation}
As follows from definition of the $R$-matrix these currents admit large $u$ expansion
\begin{equation}\label{currents-large-u-expansion}
	h(u)=1+\frac{h_{0}}{u}+\frac{h_{1}}{u^{2}}+\dots,\quad e(u)=\frac{e_{0}}{u}+\frac{e_{1}}{u^{2}}+\dots,\quad
	f(u)=\frac{f_{0}}{u}+\frac{f_{1}}{u^{2}}+\dots,\quad \psi(u)=1+\frac{\psi_{0}}{u}+\frac{\psi_{1}}{u^{2}}+\dots
\end{equation}

Using the definition \eqref{efh-def} and \eqref{psi-definition} and explicit expression for the $R$-matrix on first three levels one can prove \cite{Litvinov:2020zeq} the following relations
\begin{subequations}\label{Yangian-relation-main}
\begin{gather}
		[h(u),\psi(v)]=0,\quad[\psi(u),\psi(v)]=0,\quad[h(u),h(v)]=0,\label{hh-relation}\\
		(u-v-\epsilon_{3})h(u)e(v)=(u-v)e(v)h(u)\textcolor{blue}{-\epsilon_3 h(u) e(u)}, \label{he-relation} \\ (u-v-\epsilon_{3})f(v)h(u)=(u-v)h(u)f(v)\textcolor{blue}{-\epsilon_3 f(u)h(u)},\label{hf-relation} \\
		[e(u),f(v)]=\frac{\psi(u)-\psi(v)}{u-v}\label{ef-relation}, 
\end{gather}
as well as $ee$, $ff$ relations
\begin{multline}\label{ee-exact-relation}
		g(u-v)\Bigl[e(u)e(v)\textcolor{blue}{-\frac{e_{\includegraphics[scale=0.035]{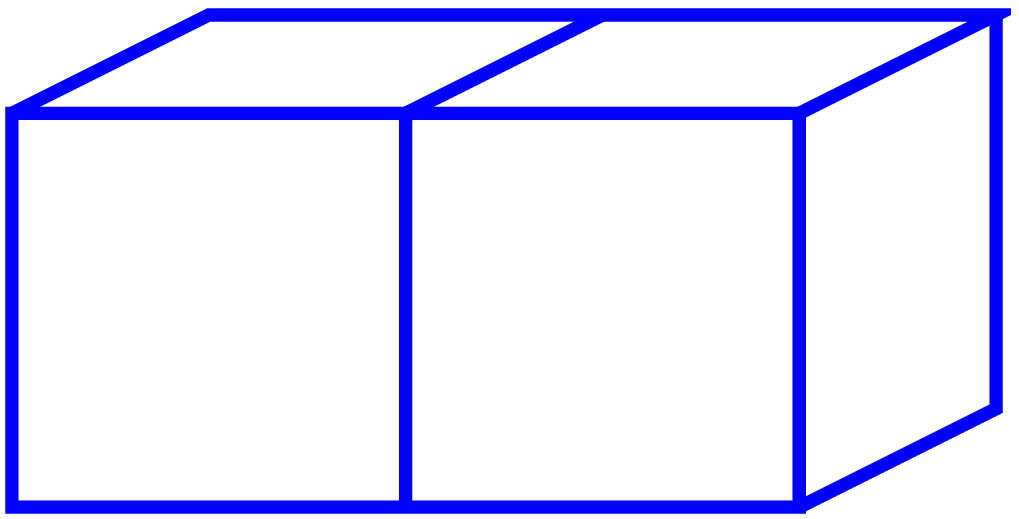}}(v)}{u-v+\epsilon_{1}}-
			\frac{e_{\includegraphics[scale=0.035]{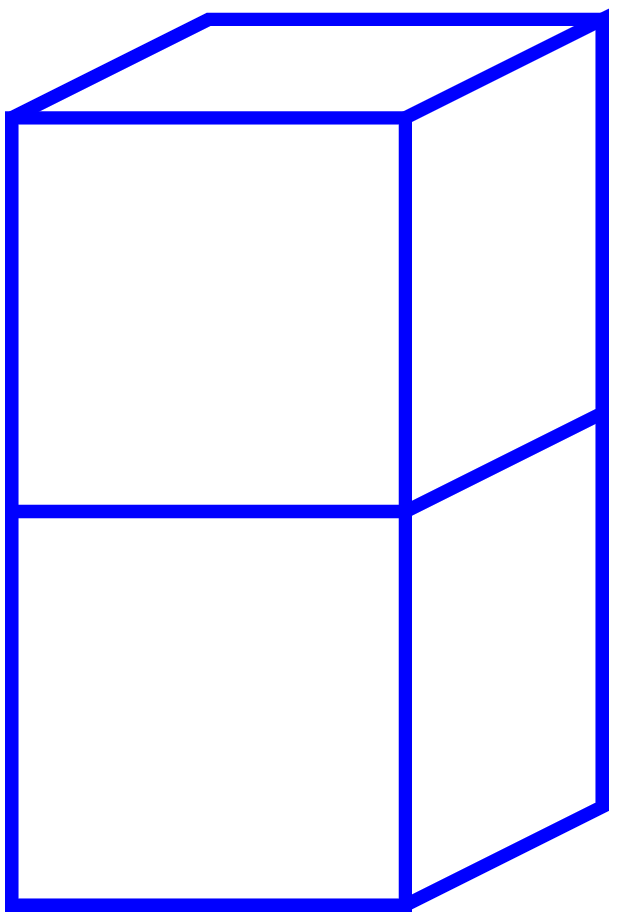}}(v)}{u-v+\epsilon_{2}}-\frac{e_{\includegraphics[scale=0.035]{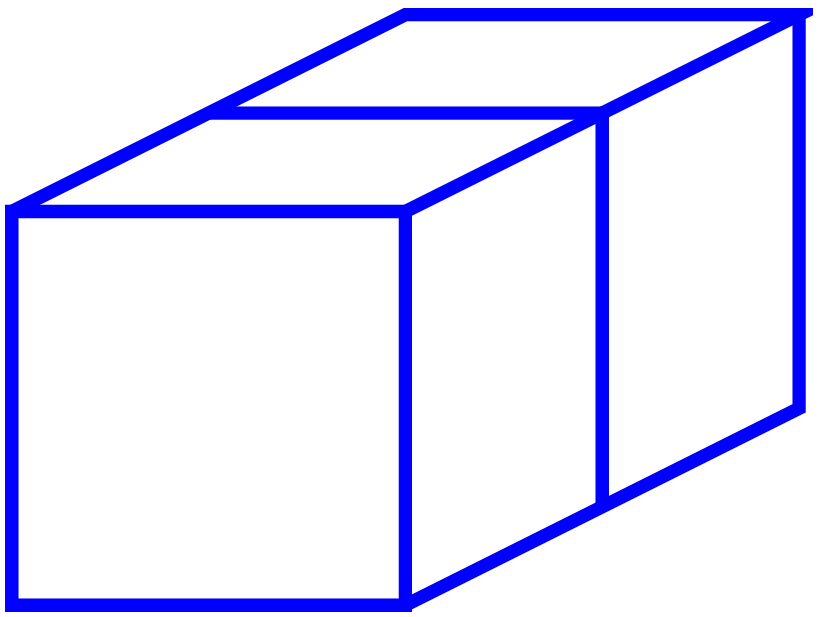}}(v)}{u-v+\epsilon_{3}} }\Bigr]=\\=
		\bar{g}(u-v)\Bigl[e(v)e(u)\textcolor{blue}{-\frac{e_{\includegraphics[scale=0.035]{partition2-blue.eps}}(u)}{u-v-\epsilon_{1}}-
			\frac{e_{\includegraphics[scale=0.035]{partition3-blue.eps}}(u)}{u-v-\epsilon_{2}}-\frac{e_{\includegraphics[scale=0.035]{partition1-blue.eps}}(u)}{u-v-\epsilon_{3}}}\Bigr], 
		\end{multline}
		\begin{multline}\label{ff-exact-relation}
		\bar{g}(u-v)\Bigl[f(u)f(v)\textcolor{blue}{-\frac{f_{\includegraphics[scale=0.035]{partition2-blue.eps}}(v)}{u-v-\epsilon_{1}}-
			\frac{f_{\includegraphics[scale=0.035]{partition3-blue.eps}}(v)}{u-v-\epsilon_{2}}-\frac{f_{\includegraphics[scale=0.035]{partition1-blue.eps}}(v)}{u-v-\epsilon_{3}}}\Bigr]=\\=
		g(u-v)\Bigl[f(v)f(u)\textcolor{blue}{-\frac{f_{\includegraphics[scale=0.035]{partition2-blue.eps}}(u)}{u-v+\epsilon_{1}}-
			\frac{f_{\includegraphics[scale=0.035]{partition3-blue.eps}}(u)}{u-v+\epsilon_{2}}-\frac{f_{\includegraphics[scale=0.035]{partition1-blue.eps}}(u)}{u-v+\epsilon_{3}}}\Bigr],
\end{multline}
$\psi e$, $\psi f$ relations
\begin{equation}\label{psi-ef-relations}
    \begin{aligned}
      &g(u-v)\psi(u)e(v)=\bar{g}(u-v)e(v)\psi(u)+\textcolor{blue}{\text{locals}},\\
      &g(u-v)f(v)\psi(u)=\bar{g}(u-v)\psi(u)f(v)+\textcolor{blue}{\text{locals}},
    \end{aligned}
\end{equation}
and Serre relations
\begin{equation}\label{Serre-relations}
\begin{gathered}    
	\sum_{\sigma\in\mathbb{S}_{3}}(u_{\sigma_{1}}-2u_{\sigma_{2}}+u_{\sigma_{3}})e(u_{\sigma_{1}})e(u_{\sigma_{2}})e(u_{\sigma_{3}})\textcolor{blue}{+\sum_{\sigma\in\mathbb{S}_{3}}
	[e(u_{\sigma_1}),e_{\includegraphics[scale=0.035]{partition1-blue.eps}}(u_{\sigma_2})+e_{\includegraphics[scale=0.035]{partition2-blue.eps}}(u_{\sigma_2})+e_{\includegraphics[scale=0.035]{partition3-blue.eps}}(u_{\sigma_2})]}=0, \\ 
	\sum_{\sigma\in\mathbb{S}_{3}}(u_{\sigma_{1}}-2u_{\sigma_{2}}+u_{\sigma_{3}})f(u_{\sigma_{1}})f(u_{\sigma_{2}})f(u_{\sigma_{3}})\textcolor{blue}{+\sum_{\sigma\in\mathbb{S}_{3}}
	[f(u_{\sigma_1}),f_{\includegraphics[scale=0.035]{partition1-blue.eps}}(u_{\sigma_2})+f_{\includegraphics[scale=0.035]{partition2-blue.eps}}(u_{\sigma_2})+f_{\includegraphics[scale=0.035]{partition3-blue.eps}}(u_{\sigma_2})]}=0. 
\end{gathered}
\end{equation}
\end{subequations}
In the relations above we have used the following notations
\begin{equation}
 g(x)\overset{\text{def}}{=}(x+\epsilon_{1})(x+\epsilon_{2})(x+\epsilon_{3}),\quad
 \bar{g}(x)\overset{\text{def}}{=}(x-\epsilon_{1})(x-\epsilon_{2})(x-\epsilon_{3}).
\end{equation}

We note that the terms shown by \textcolor{blue}{blue} in \eqref{he-relation}-\eqref{psi-ef-relations} depend only on one parameter either $u$ or $v$ (in \eqref{psi-ef-relations} these terms are so complicated, that we do no write them explicitly) and in \eqref{ff-exact-relation} they depend only on two parameters instead of one.  We call such terms \emph{local}, the main idea is that they always can be omitted in actual computations, as only interested in relations between modes of currents they always will stand inside some contour integral, and the integration contour always can be chosen in a way to exclude local terms. 

Now let us describe the inverse map from the Borel sub-algebra of $\textrm{RLL}$ algebra to the currents. We introduce the modes $U_n$ of $W^{(1)}(z)$ current
\begin{align}
 &\langle\varnothing|\mathcal{L}(u) \  a^{(0)}_{-n}|\varnothing\rangle=\frac{U_n}{u}+O\left(\frac{1}{u^2}\right), \quad  n>0 \label{Jn+}\\
 &\langle\varnothing|a^{(0)}_{n}\ \mathcal{L}(u) |\varnothing\rangle=\frac{U_{-n}}{u}+O\left(\frac{1}{u^2}\right), \quad  n>0 \label{Jn-}
\end{align}
It is clear from the $\textrm{RLL}$ relation that the  $R-$matrix commutes with the $W^{(1)}$ current:
\begin{equation}
(a_n^{(0)}+U_n)\mathcal{R}_{0,v}=\mathcal{R}_{0,v}(a_n^{(0)}+U_n)
\end{equation}
Taking the matrix element over the auxiliary space $\langle\varnothing|\dots |\mu \rangle$  for positive $n$ we will get:
\begin{equation}\label{Ad}
[\mathcal{L}_{\boldsymbol{\mu},\varnothing}(u),U_n]=\mathcal{L}_{\boldsymbol{\mu}+n,\varnothing}(u), 
\end{equation}
where $\langle\boldsymbol{\mu}+n| \overset{\text{def}}{=} \langle \boldsymbol{\mu}|a_n$.
	
It is also clear, that $U_n$ for $n>0$ belongs to the subalgebra $\mathfrak{n}^+$. Indeed, explicit calculation of the large $u$ limit of $\mathcal{R}(u)$ (see \cite{Litvinov:2020zeq} for the details) shows that:
\begin{align} \label{Jef}
 U_1=f_0 &\quad U_{-1}=e_0,\\
	U_{k+1}=-k[f_1,U_k] &\quad U_{k-1}=-k[e_1,U_{k}].
\end{align}
Then we get:
\begin{equation}\label{JSh}
 U_{k}^{\boldsymbol{x}}=\oint\dots\oint g_k(\boldsymbol{z})f(z_1)...f(z_k)d\boldsymbol{z}\quad\text{with}\quad
 g_{k+1}(\boldsymbol{z})=-k\Big(z_1g_k(z_2,\dots ,z_{k+1})-g_n(z_1,\dots ,z_k)z_{k+1}\Big),
\end{equation}
and
\begin{equation}
g_{k}(\boldsymbol{z})=(-1)^{k-1} (k-1)! \prod\limits_i z_i\left(\sum (-1)^iC_k^i z^{-1}_i\right),
\end{equation}
where $C_n^i$ are the binomial coefficients. 

Finally using \eqref{Ad} we may express $\mathcal{L}_{\boldsymbol{\lambda},\varnothing}(u)$ as a multiple commutator of $\mathcal{L}_{\varnothing,\varnothing}(u)=h(u)$ and modes of $f(z)$ currents, or equivalently as contour integral
\begin{equation}\label{LAsSh}
\mathcal{L}_{\boldsymbol{\lambda},\varnothing}(u)=\frac{1}{(2\pi i)^{|{\lambda}|}}\oint\dots\oint F_{\boldsymbol{\lambda}}(\boldsymbol{z}|u)\,h(u)f(z_{|\boldsymbol{\lambda}|})\dots f(z_1)dz_1\dots, dz_{|\boldsymbol{\lambda}|}
\end{equation} 
with some explicit function $F_{{\lambda}}(\boldsymbol{z}|u)$.
\subsection{Antipode}
As we will see there is an important operation: the reflection of the boson $\varphi(x)\to -\varphi(x)$. Using it we define the antipode of $L$-operator:
\begin{gather}\label{barLAsSh}
    (\mathcal{L}_{\boldsymbol{\mu},\boldsymbol{\nu}}(u))^{a}\overset{\text{def}}{=}\bar{\mathcal{L}}_{\boldsymbol{\mu},\boldsymbol{\nu}}(u)=(-1)^{l(\boldsymbol{\mu})+l(\boldsymbol{\nu})}\mathcal{L}(-u)_{\boldsymbol{\nu},\boldsymbol{\mu}}, \\
    \overline{\mathcal{L}(u)\mathcal{L}(v)}\overset{\text{def}}{=}\bar{\mathcal{L}}(v)\bar{\mathcal{L}}(u).
\end{gather}
Here $l(\boldsymbol{\mu})$ is the number of rows in Young diagram $\boldsymbol{\mu}$\footnote{Note that if we thing of the diagram as of the bosonic state: $|\lambda\rangle =\prod\limits_{i=1}^{l(\lambda)} a_{-\lambda_i}|\varnothing\rangle$, then multiplication by $(-1)^{\lambda}$ is nothing but the reflection of the bosons $a_{-n}\to -a_{-n}$.}.

It is convenient to write the conjugated $L$ operator as follows:
\begin{equation} \label{barL}
\bar{\mathcal{L}}_{\boldsymbol{\lambda},\varnothing}(u)=\frac{1}{(2\pi i)^{|\boldsymbol{\lambda}|}}\oint\dots\oint F_{\boldsymbol{\lambda}}(\boldsymbol{z}|u)\,f(-\epsilon_3-z_{|{\lambda}|})\dots f(-\epsilon_3-z_1) h(-u) dz_1\dots dz_{|{\lambda}|}
\end{equation} 
\section{Off-shell Bethe vector}\label{off-shell-introduction}
In order to construct the off-shell Bethe vector we consider the tensor product of $n+N$ Fock spaces
\begin{equation}\label{total-Fock}
\underbrace{\mathcal{F}_{u_{n}}\otimes\dots\otimes\mathcal{F}_{u_{1}}}_{\text{quantum space}}\otimes
\underbrace{\mathcal{F}_{x_{1}}\otimes\dots\otimes\mathcal{F}_{x_{N}}}_{\text{auxiliary space}}=\mathcal{F}_{\boldsymbol{u}}\otimes F_{\boldsymbol{x}}
\end{equation}
generated from the vacuum state
\begin{equation}
|\boldsymbol{\varnothing}\rangle_{\boldsymbol{u}}\otimes |\boldsymbol{\varnothing}\rangle_{\boldsymbol{x}}=|u_{n}\rangle
\otimes\dots\otimes|u_{1}\rangle\otimes|x_{1}\rangle\otimes\dots\otimes|x_{N}\rangle.
\end{equation}
In order not to confuse between the auxiliary and quantum Fock spaces, we will label Fock space not by it's index, but by it's spectral parameter. So that the $R$-matrix between two Fock spaces will read as $\mathcal{R}_{u_i,u_j}$ while the $R$-matrix between two auxiliary spaces as $\mathcal{R}_{x_i,x_j}$.

As usual let us introduce $\mathcal{L}_i(u_i)$ operators:
\begin{equation}
\mathcal{L}_i(u_i)=\mathcal{R}_{u_i,\boldsymbol{x}}=\mathcal{R}_{u_i,x_1}\dots\mathcal{R}_{u_i,x_N}, \qquad
\mathcal{L}_{\boldsymbol{u}}=\mathcal{L}_n(u_n)\dots \mathcal{L}_1(u_1).
\end{equation}
It is also convenient to define opposite $\bar{\mathcal{L}}$ operators:
\begin{equation}
  \bar{\mathcal{L}}_i(u_i)=\mathcal{R}_{\bar{u_i},\boldsymbol{x}}=\mathcal{R}_{\bar{u_i},x_N}\dots\mathcal{R}_{\bar{u_i},x_1},\qquad
\bar{\mathcal{L}}_{\boldsymbol{u}}=\bar{\mathcal{L}}_1(u_1)\dots \bar{\mathcal{L}}_n(u_n) 
\end{equation}
\subsection{K operators}
In the previous section we have defined the $K$-matrix acting on the single Fock space \eqref{definition-KR}. It is useful to extend its action to the tensor product of quantum and auxiliary Fock spaces. Let us define
\begin{equation} \label{Kvx1}
\mathcal{K}_{\boldsymbol{u}|y}\overset{\text{def}}{=}
\mathcal{R}_{\overline{\boldsymbol{u}},y}\mathcal{K}_{y}\mathcal{R}_{\boldsymbol{u},y},
\end{equation}
where
\begin{equation}
\mathcal{R}_{\boldsymbol{u},y}=\mathcal{R}_{u_n,y}\dots \mathcal{R}_{u_1,y} \, , \quad 
\mathcal{R}_{v,\boldsymbol{x}}=\mathcal{R}_{v,x_N}\dots \mathcal{R}_{v,x_1}
\end{equation}
and $\mathcal{K}_{y}$ is the operator defined in \eqref{definition-KR}. The definition \eqref{Kvx1} is the direct analog of $L$-operator to the boundary case. This definition can be conveniently illustrated with the following picture
\begin{equation*}
\psfrag{k}{$\mathcal{K}_{\boldsymbol{u}|y}=$}
\psfrag{0}{$\scriptscriptstyle{y}$}
\psfrag{1}{$\scriptscriptstyle{u_1}$}
\psfrag{2}{$\scriptscriptstyle{u_2}$}
\psfrag{3}{$\scriptscriptstyle{u_3}$}
\psfrag{4}{$\scriptscriptstyle{x_1}$}
\psfrag{5}{$\scriptscriptstyle{x_2}$}
\psfrag{6}{$\scriptscriptstyle{x_3}$}
\psfrag{7}{$\scriptscriptstyle{u_{n-1}}$}
\psfrag{8}{$\scriptscriptstyle{u_n}$}
\includegraphics[width=.5\textwidth]{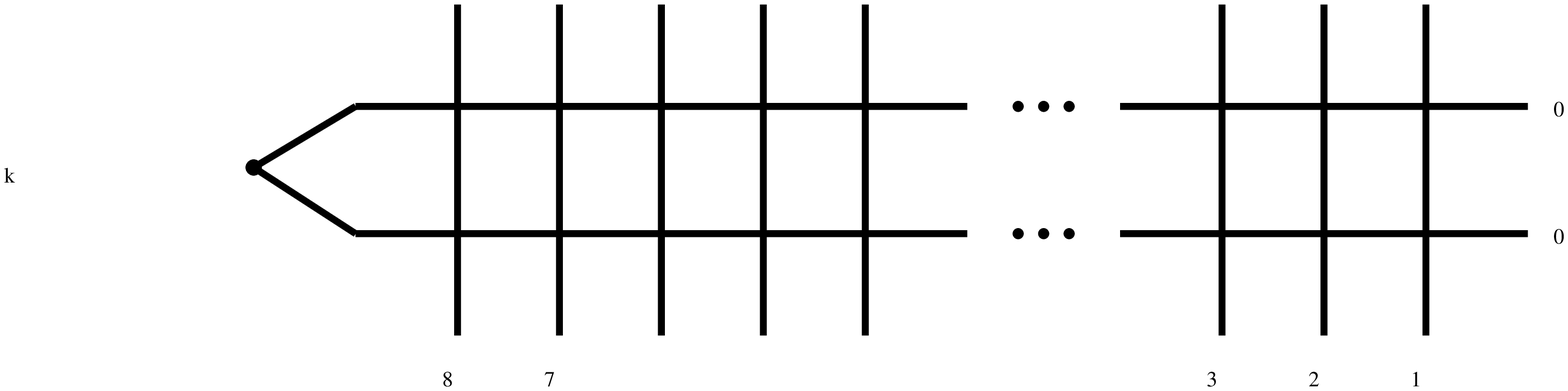}
\end{equation*}
We note that $\mathcal{K}_{\boldsymbol{u}|x_1}$ still enjoys $\textrm{KRKR}$ equation \eqref{reflection-equation}
\begin{equation}
    \mathcal{R}_{x_1,x_2}\mathcal{K}_{\boldsymbol{u}|x_1}\mathcal{R}_{x_1,\bar{x}_2}\mathcal{K}_{\boldsymbol{u}|x_2}=\mathcal{K}_{\boldsymbol{u}|x_2}\mathcal{R}_{x_1,\bar{x}_2}\mathcal{K}_{\boldsymbol{u}|x_1}\mathcal{R}_{x_1,x_2}.
\end{equation}
Now let us extend the action of our $K$-operator to the full auxiliary space $\mathcal{F}_{\boldsymbol{x}}$. The most convenient way to do it is by recurrent formula
\begin{equation}
\mathcal{K}_{\boldsymbol{u}|y,\boldsymbol{x}}=\mathcal{K}_{\boldsymbol{u}|\boldsymbol{x}}\mathcal{R}_{\bar{y},\boldsymbol{x}}\mathcal{K}_y.
\end{equation}
Here $\mathcal{K}_{\boldsymbol{u}|y}$ is the operator defined in \eqref{Kvx1} acting on a tensor product $\mathcal{F}_{\boldsymbol{u}}\otimes \mathcal{F}_y$, while $\mathcal{K}_{\boldsymbol{u}|\boldsymbol{x}}$ acts on a tensor product of $F_{\boldsymbol{u}}\otimes F_{\boldsymbol{x}}$.
The last formula can be illustrated by the following picture (here we consider for simplicity the case of $N=3$)
\begin{figure}[h!]
\centering
\psfrag{0}{$\scriptscriptstyle{}$}
\psfrag{1}{$\scriptscriptstyle{u_1}$}
\psfrag{2}{$\scriptscriptstyle{u_2}$}
\psfrag{3}{$\scriptscriptstyle{u_3}$}
\psfrag{4}{$\scriptscriptstyle{x_1}$}
\psfrag{5}{$\scriptscriptstyle{x_2}$}
\psfrag{6}{$\scriptscriptstyle{x_3}$}
\psfrag{7}{$\scriptscriptstyle{u_{n-1}}$}
\psfrag{8}{$\scriptscriptstyle{u_n}$}
\psfrag{p}{$\mathcal{K}_{\boldsymbol{u}|\boldsymbol{x}}=$}
\includegraphics[width=.7\textwidth]{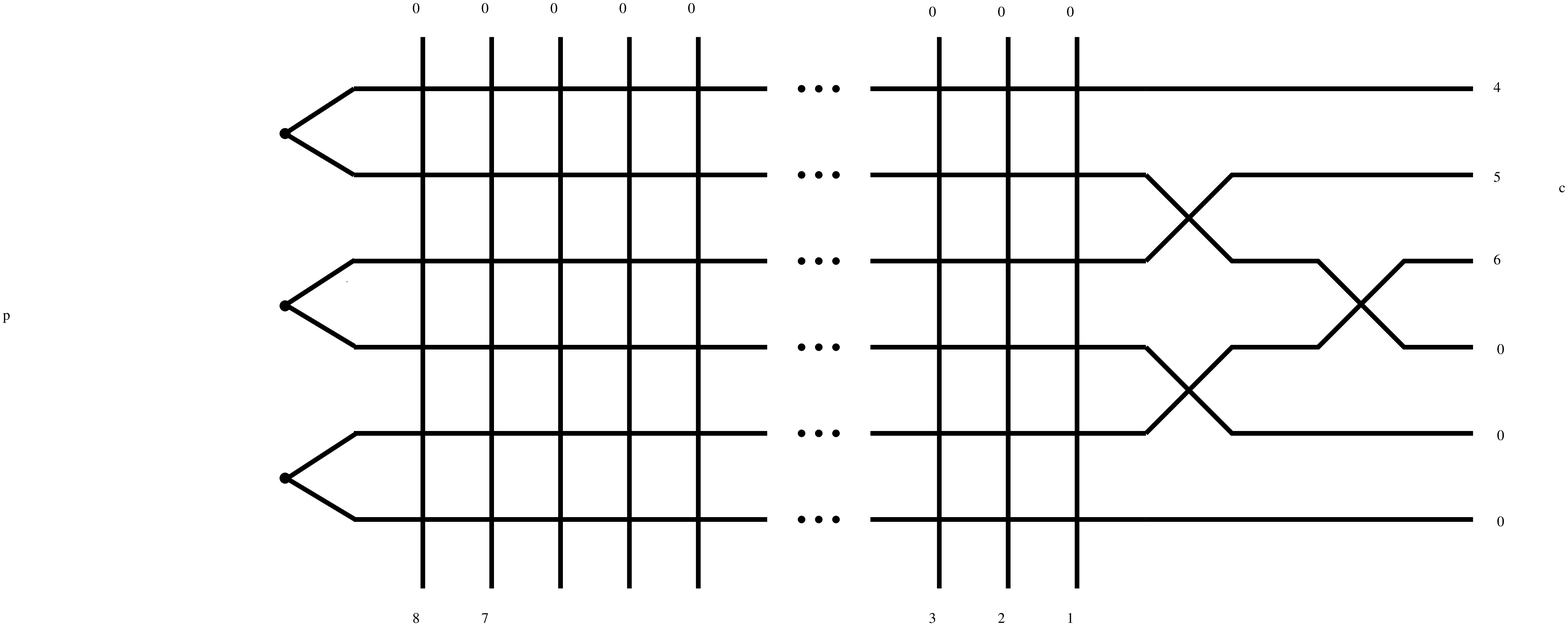}
\caption{Explicitly this $\mathcal{K}$ operator reads as $\mathcal{K}_{\boldsymbol{u}|x_3,x_2,x_1}=\mathcal{K}_{\boldsymbol{u}|x_3}\mathcal{R}_{\bar{x}_2,x_3}\mathcal{K}_{\boldsymbol{u}|x_2}\mathcal{R}_{\bar{x}_1,x_3}\mathcal{R}_{\bar{x}_1,x_2}\mathcal{K}_{\boldsymbol{u}|x_1}$}
\end{figure}

Finally our definition of $\mathcal{K}_{{\boldsymbol{v}|\boldsymbol{x}}}$ may be summarised in two operations which increase the number of quantum and auxiliary Fock spaces:
\begin{equation}
\Delta^q(\mathcal{K}_{\boldsymbol{u}|\boldsymbol{x}})=\mathcal{K}_{v,\boldsymbol{u}|\boldsymbol{x}}=\bar{\mathcal{L}}_v\mathcal{K}_{{\boldsymbol{u}|\boldsymbol{x}}}\mathcal{L}_v,\qquad
\Delta^a(\mathcal{K}_{\boldsymbol{u}|\boldsymbol{x}})=\mathcal{K}_{{\boldsymbol{u}|y,\boldsymbol{x}}}=\mathcal{K}_{\boldsymbol{u}|\boldsymbol{x}}\mathcal{R}_{\bar{y},\boldsymbol{x}}\mathcal{K}_y.
\end{equation}

Using the Yang-Baxter equation and $\textrm{KRKR}$ relation, one can show that two operators actually commute $\Delta^q(\Delta^a(\mathcal{K}_{{\boldsymbol{v}|\boldsymbol{x}}}))=\Delta^a(\Delta^q(\mathcal{K}_{{\boldsymbol{v}|\boldsymbol{x}}}))$. This property may be illustrated by the following picture
\begin{equation*}
\psfrag{e}{$=$}
\psfrag{0}{$\scriptscriptstyle{y}$}
\psfrag{1}{$\scriptscriptstyle{u_1}$}
\psfrag{2}{$\scriptscriptstyle{u_2}$}
\psfrag{3}{$\scriptscriptstyle{u_3}$}
\psfrag{4}{$\scriptscriptstyle{x_1}$}
\psfrag{5}{$\scriptscriptstyle{x_2}$}
\psfrag{6}{$\scriptscriptstyle{x_3}$}
\psfrag{7}{$\scriptscriptstyle{u_{n-1}}$}
\psfrag{8}{$\scriptscriptstyle{u_n}$}
\includegraphics[width=.9\textwidth]{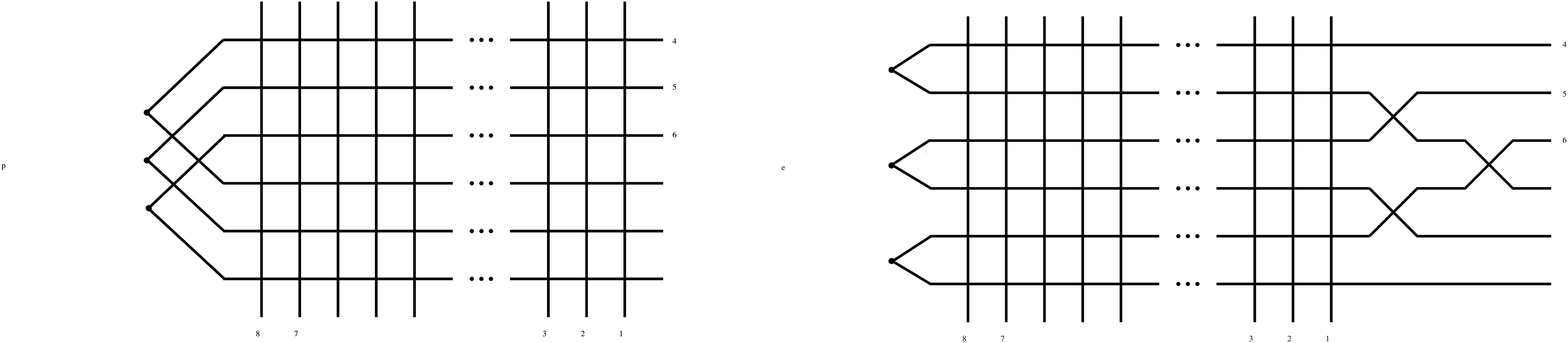}
\end{equation*}
\subsection{Off-shell Bethe vector}
Now we are ready to introduce the off-shell Bethe vector 
\begin{equation}\label{definition-off-shell-bethe-vector}
|B(\boldsymbol{x})\rangle=_x\! \! \langle \varnothing|\bar{\mathcal{L}}_{{\boldsymbol{v}}}\mathcal{K}_{\boldsymbol{x}}{L}_{{\boldsymbol{v}}}|\varnothing\rangle_v |\chi\rangle_x=_x\! \! \langle \varnothing|\mathcal{K}_{{\boldsymbol{v}}|\boldsymbol{x}}|\varnothing\rangle_v |\chi\rangle_x
\end{equation}
here $|\chi\rangle_x$ is a special state in auxiliary Fock space: $|\chi\rangle_x \in \mathcal{F}_{\boldsymbol{x}}$. It has  grading $N$ with respect to the standard grading operator. The vector $|B(\boldsymbol{x})\rangle$ can be represented by the following picture
\footnote{This formula differs from the one provided by Sklyanin \cite{Sklyanin:1988yz} in $\mathfrak{sl}(2)$ case. In his approach $\mathcal{K}_{\boldsymbol{u}|\boldsymbol{x}}$ operator is a product of single space operators $\prod\limits_i 
\mathcal{K}_{\boldsymbol{u}|x_i}$. It can be shown, that for $\mathfrak{sl}(2)$ case these two approaches coincide. For example on level $3$ we have: 
\begin{equation*}
|B(\boldsymbol{x})\rangle= _x\!\langle\downarrow\downarrow\downarrow|\mathcal{K}_{\boldsymbol{u}|x_3,x_2,x_1}|\uparrow\uparrow\uparrow\rangle_x\otimes|\boldsymbol{\downarrow}\rangle_u=_x\!\langle\downarrow\downarrow\downarrow|\mathcal{K}_{\boldsymbol{u}|x_3}\mathcal{R}_{\bar{x}_2,x_3}\mathcal{K}_{\boldsymbol{u}|x_2}\mathcal{R}_{\bar{x}_1,x_3}\mathcal{R}_{\bar{x}_1,x_2}\mathcal{K}_{\boldsymbol{u}|x_1}|\uparrow\uparrow\uparrow\rangle_x\otimes|\boldsymbol{\downarrow}\rangle_u.
\end{equation*}
In the case of $\mathfrak{sl}(2)$, $\textrm{R}$-matrices between the auxiliary spaces may be omitted, and we reproduce Sklyanin's formula:
\begin{equation*}
   |B(\boldsymbol{x})\rangle \overset{\text{for} \ \mathfrak{sl}(2)}{=}_x\!\langle\downarrow\downarrow\downarrow|\mathcal{K}_{\boldsymbol{u}|x_3}\mathcal{K}_{\boldsymbol{u}|x_2}\mathcal{K}_{\boldsymbol{u}|x_1}|\uparrow\uparrow\uparrow\rangle_x\otimes|\boldsymbol{\downarrow}\rangle_u=\langle\downarrow|\mathcal{K}_{\boldsymbol{u}|x_3}|\uparrow\rangle\langle\downarrow|\mathcal{K}_{\boldsymbol{u}|x_2}|\uparrow\rangle \langle\downarrow|\mathcal{K}_{\boldsymbol{u}|x_1}|\uparrow\rangle |\boldsymbol{\downarrow}\rangle_u.
\end{equation*}
}
\begin{equation}\label{B-picture}
   \psfrag{cc}{\resizebox{1cm}{!}{$|\varnothing\rangle$}}
   \psfrag{0}{$\scriptscriptstyle{\boldsymbol{\varnothing}}$}
   \psfrag{1}{$\scriptscriptstyle{u_1}$}
   \psfrag{2}{$\scriptscriptstyle{u_2}$}
   \psfrag{3}{$\scriptscriptstyle{u_3}$}
   \psfrag{4}{$\scriptscriptstyle{x_1}$}
   \psfrag{5}{$\scriptscriptstyle{x_2}$}
   \psfrag{6}{$\scriptscriptstyle{x_3}$}
   \psfrag{7}{$\scriptscriptstyle{u_{n-1}}$}
   \psfrag{8}{$\scriptscriptstyle{u_n}$}
   \psfrag{p}{$|B(\boldsymbol{x})\rangle=$}
   \psfrag{c}{\resizebox{1.1cm}{!}{$|\chi\rangle_{\scriptscriptstyle{\boldsymbol{x}}}$}}
   \includegraphics[width=.55\textwidth]{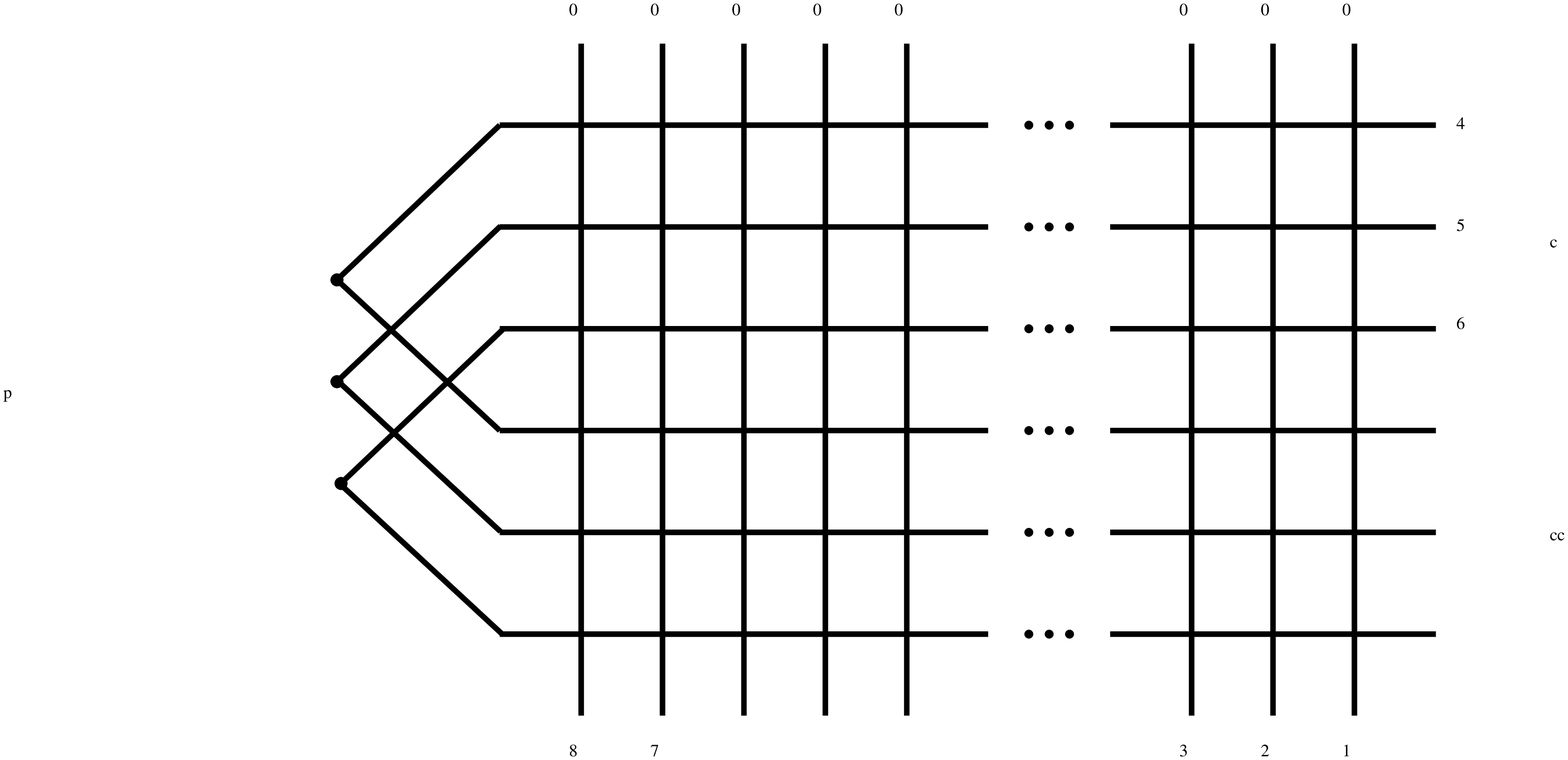}
\end{equation}
We will introduce the explicit form of $|\chi\rangle_x$ in the next section \eqref{chi-definition}, for now let us just announce the main property of Bethe vector.
\subsection{Bethe Ansatz equations, eigenvalues of KZ IOMs.}
We note that the Yang-Baxter and $\textrm{KRKR}$ relation \eqref{reflection-equation} provide the nice intertwining property of the off-shell Bethe vector with the $\mathcal{T}_i^+$ operator \eqref{KZ_oper}
\begin{equation}
    \mathcal{T}_i^+|B(\boldsymbol{x})\rangle=|B(\boldsymbol{x})\rangle\Big|_{\varphi_i\to-\varphi_i}
\end{equation}
In the next section we will prove that under the Bethe ansatz equations
\begin{equation}\label{BAE}
\begin{gathered}
   r^{\alpha}(x_i)r^{\beta}(x_i)A(x_i)A^{-1}(-x_i)\prod_{j\neq i}G(x_i-x_j)G^{-1}(-x_i-x_j)=1,
\\
G(x)=\frac{(x-\epsilon_1)(x-\epsilon_2)(x-\epsilon_3)}{(x+\epsilon_1)(x+\epsilon_2)(x+\epsilon_3)},\qquad
  A(x)=\prod_{k=1}^n\frac{x-u_k+\frac{\epsilon_3}{2}}{x-u_k-\frac{\epsilon_3}{2}},\quad r^{\alpha}(x)=-\frac{x+\epsilon_{\alpha}/2}{x-\epsilon_{\alpha}/2}.
\end{gathered}
\end{equation}
the off-shell Bethe vector with shifted $x$ parameters $|B(\boldsymbol{x}-\frac{\epsilon_3}{2})\rangle$ becomes an eigenvector of \textrm{KZ} IOMs $\mathcal{I}_i^{\textrm{KZ}}$ \eqref{KZ_oper}:
\begin{equation}\label{BAE-eignvalue}
\mathcal{I}_i^{\textrm{KZ}} |B(\boldsymbol{x}-\frac{\epsilon_3}{2})\rangle\overset{\text{BAE}(\boldsymbol{x})=1}{=}\prod\limits_a \frac{(u_i+\frac{\epsilon_3}{2})^2-x_a^2}{(u_i-\frac{\epsilon_3}{2})^2-x_a^2}|B(\boldsymbol{x}-\frac{\epsilon_3}{2})\rangle.
\end{equation}
Equations \eqref{BAE} and \eqref{BAE-eignvalue} together with the explicit form of off-shell Bethe vector \eqref{definition-off-shell-bethe-vector} are the main results of our paper.
\section{Diagonalization of KZ integral}\label{BA-proof}
Let us revise the formula for the off-shell Bethe vector \eqref{definition-off-shell-bethe-vector}.  One observes that the definition of \eqref{definition-off-shell-bethe-vector} (as especially seen from the picture \eqref{B-picture}) suggests that $|B(\boldsymbol{x})\rangle$ can be interpreted as a product of some $L-$operators $\mathfrak{L}(u_n)\dots\mathfrak{L}(u_1)$ sandwiched between bra and ket states $\langle\mathcal{K}_{\boldsymbol{x}}|$ and $\Big|\begin{matrix} \chi \\ \varnothing \end{matrix}\Big\rangle_x$ 
\begin{equation}\nonumber
 \psfrag{cc}{\resizebox{1cm}{!}{$|\varnothing\rangle$}}
   \psfrag{0}{$\scriptscriptstyle{\varnothing}$}
   \psfrag{1}{$\scriptscriptstyle{u_1}$}
   \psfrag{2}{$\scriptscriptstyle{u_2}$}
   \psfrag{3}{$\scriptscriptstyle{u_3}$}
   \psfrag{4}{$\scriptscriptstyle{x_1}$}
   \psfrag{5}{$\scriptscriptstyle{x_2}$}
   \psfrag{6}{$\scriptscriptstyle{x_3}$}
   \psfrag{7}{$\scriptscriptstyle{u_{n-1}}$}
   \psfrag{8}{$\scriptscriptstyle{u_n}$}
   \psfrag{p}{$|B(\boldsymbol{x})\rangle=$}
   \psfrag{K}{$\langle\mathcal{K}_{\boldsymbol{x}}|$}
   \psfrag{L}{$\! \! \! \mathfrak{L}(u_2)$}
   \psfrag{c}{\resizebox{1.1cm}{!}{$|\chi\rangle_{\scriptscriptstyle{\boldsymbol{x}}}$}}
    \psfrag{X}{$\Big|\begin{matrix} \chi \\ \varnothing \end{matrix}\Big\rangle_x$}
   \includegraphics[width=.55\textwidth]{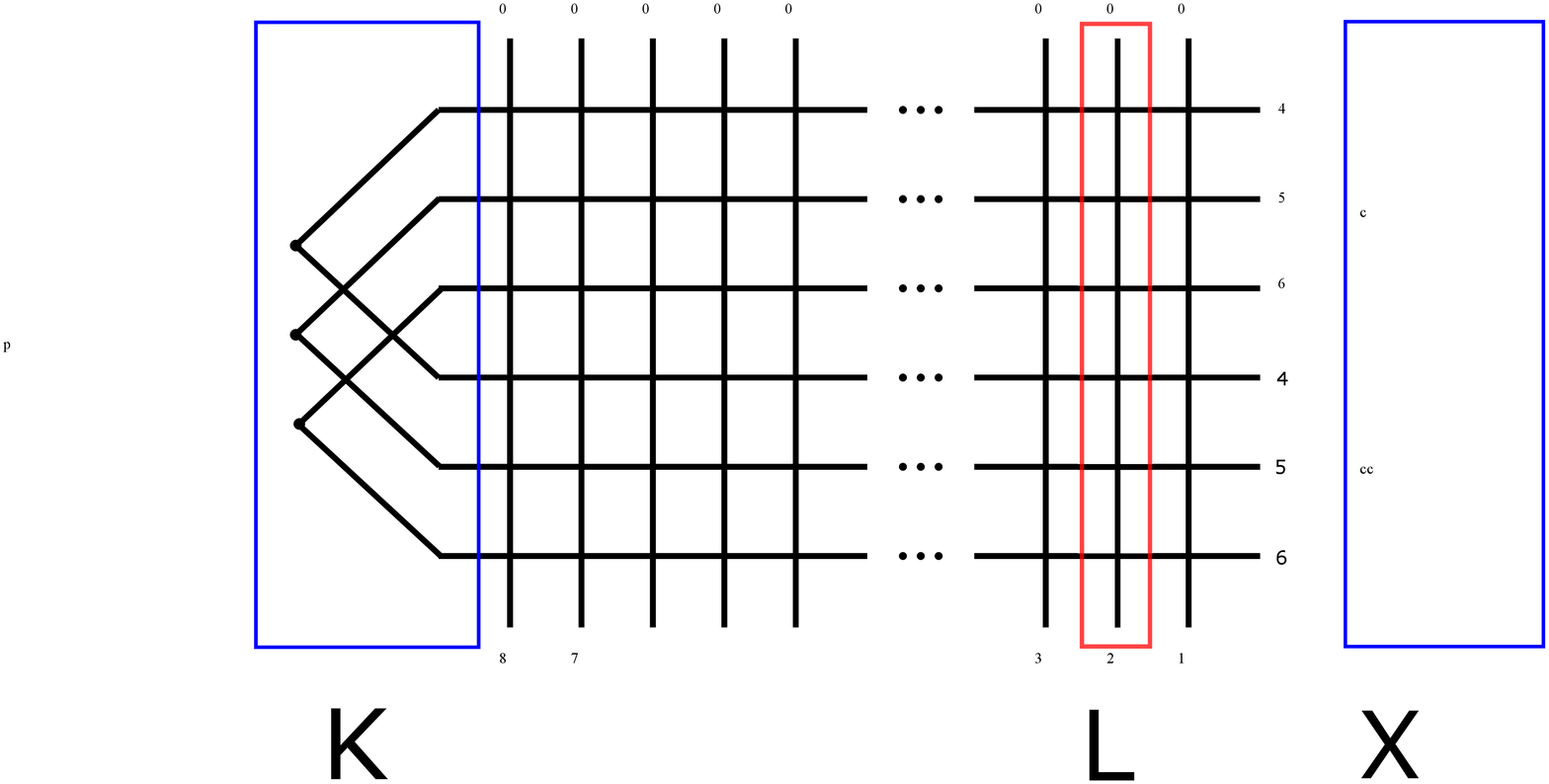}
\end{equation}
This observation can be formalized as follows. Let us define  $\mathfrak{L}(u)$ operator by the picture
\begin{equation*}
\psfrag{1}{$\mathcal{F}_{x_1}$}
\psfrag{2}{$\mathcal{F}_{x_2}$}
\psfrag{3}{$\mathcal{F}_{x_3}$}
\psfrag{4}{$\mathcal{F}^{\star}_{x_1}$}
\psfrag{5}{$\mathcal{F}^{\star}_{x_2}$}
\psfrag{6}{$\mathcal{F}^{\star}_{x_3}$}
   \psfrag{l}{$\lambda$}
   \psfrag{m}{$\mu$}
   \psfrag{L}{$\mathfrak{L}(u)_{\lambda,\mu}=$}
   \includegraphics[width=0.5\textwidth]{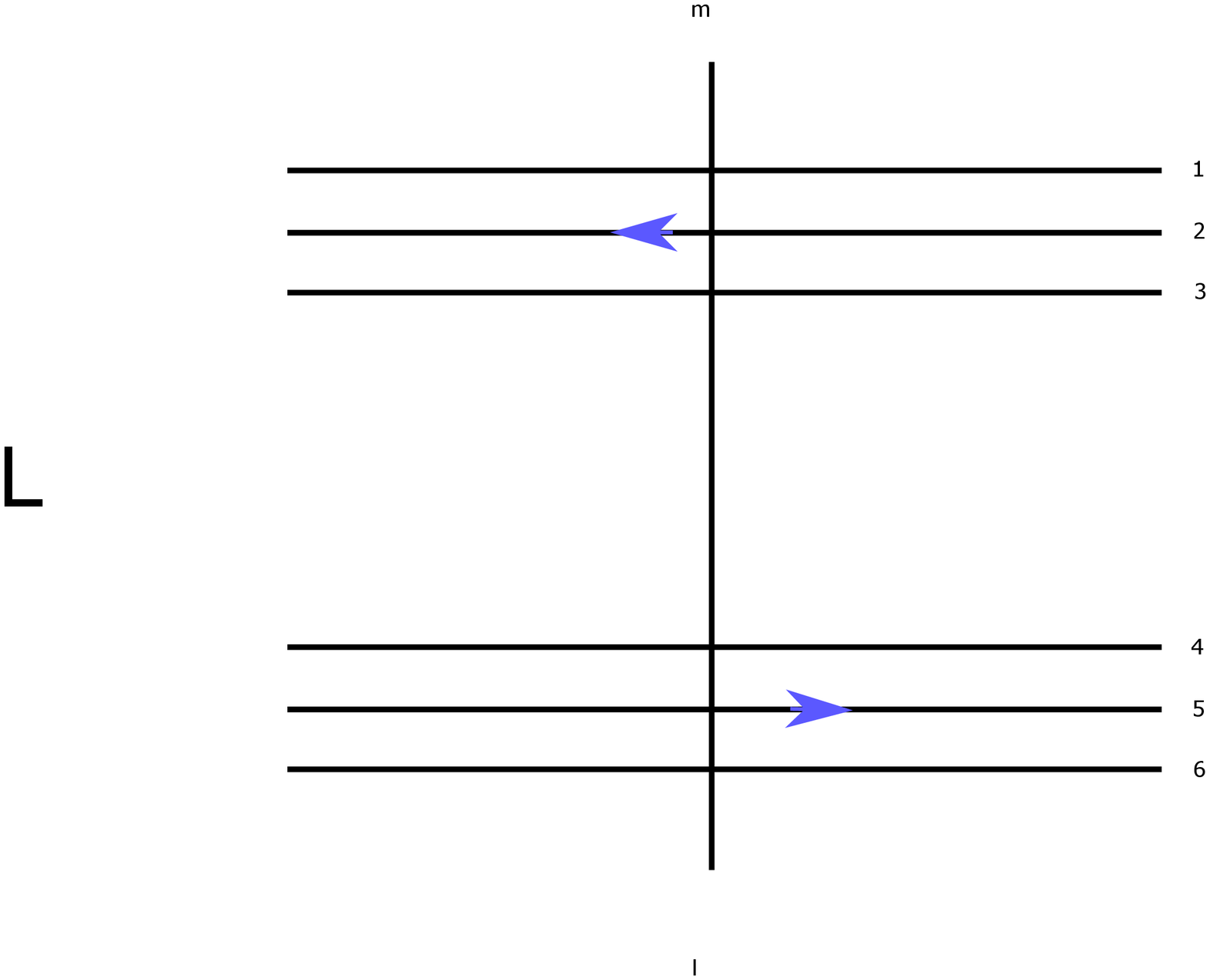}
\end{equation*}
By definition it acts in the tensor product of Fock module and it's dual $\mathcal{F}_{\boldsymbol{x}}\otimes \mathcal{F}_{\boldsymbol{x}}^{\star}$ and equals to the infinite sum
\begin{equation} \label{Loper}
\mathfrak{L}(u)_{\boldsymbol{\lambda},\boldsymbol{\mu}}=\sum\limits_{\rho}\mathcal{L}(u)_{\boldsymbol{\lambda},\boldsymbol{\rho}}\otimes \bar{\mathcal{L}}(u)_{\boldsymbol{\rho},\boldsymbol{\mu}}
\end{equation}
It is though clear that $\mathfrak{L}(u)$-operators, still enjoys $\textrm{RLL}$ algebra
\begin{equation}\label{RLL}
\mathcal{R}(u_1-u_2)\mathfrak{L}(u_1) \mathfrak{L}(u_2)=\mathfrak{L}(u_2) \mathfrak{L}(u_1)\mathcal{R}(u_1-u_2)
\end{equation}
Using this equation we may define the currents in complete analogy with $\textrm{A}$ case with exactly the same commutation relations \eqref{Yangian-relation-main}:
\begin{equation}\label{efh-def-new}
 \mathfrak{h}(u)\overset{\text{def}}{=}\mathfrak{L}_{\scriptscriptstyle{\varnothing,\varnothing}}(u),\qquad
 \mathfrak{e}(u)\overset{\text{def}}{=}\mathfrak{h}^{-1}(u)\mathfrak{L}_{\scriptscriptstyle{\varnothing,\Box}}(u)\quad\text{and}\quad
 \mathfrak{f}(u)\overset{\text{def}}{=}\mathfrak{L}_{\scriptscriptstyle{\Box,\varnothing}}(u) \mathfrak{h}^{-1}(u).
\end{equation}

Exploiting this picture further, we can consider $K-$operator $\mathcal{K}_{\boldsymbol{x}}$ as a bra vector $\langle\mathcal{K}|$ acting from $\mathcal{F}_{\boldsymbol{x}}\otimes \mathcal{F}^{\star}_{\boldsymbol{x}}$ to $\mathbb{C}$. We will denote vectors from $\mathcal{F}_{\boldsymbol{x}}\otimes \mathcal{F}^{\star}_{\boldsymbol{x}}$ by a two rows objects $\Big|\begin{matrix}\boldsymbol{\lambda} \\ \boldsymbol{\mu}
\end{matrix} \Big\rangle$ , where $\boldsymbol{\lambda} \in \mathcal{F}_{\boldsymbol{x}}$ , $\boldsymbol{\mu} \in \mathcal{F}^{\star}_{\boldsymbol{x}}$. It allows to rewrite Bethe vector \eqref{definition-off-shell-bethe-vector} as follows:
\begin{equation}\label{Bethe-vector-rotated-picture}
|B^{\alpha,\boldsymbol{u}}(\boldsymbol{x})\rangle=_{\boldsymbol{x}}\! \! \langle\varnothing|\bar{\mathcal{L}}_1\dots\bar{\mathcal{L}}_n \mathcal{K}^{\alpha}_{\boldsymbol{x}} \mathcal{L}_n \dots \mathcal{L}_1 |\chi\rangle_{\boldsymbol{x}} |\varnothing\rangle_{\boldsymbol{u}} \equiv\,_{\boldsymbol{x}}\langle \mathcal{K}^{\alpha}|\mathfrak{L}_n\dots \mathfrak{L}_1 \Big|\begin{matrix} \chi \\ \varnothing \end{matrix}\Big\rangle_{\boldsymbol{x}} |\varnothing\rangle_{\boldsymbol{u}}.
\end{equation}
The benefit of this approach is that the structure of the Bethe vector may be analysed by the representation theory of the Affine Yangian in $\mathcal{F}_{\boldsymbol{x}}\otimes \mathcal{F}_{\boldsymbol{x}}^{\star}$. We call the corresponding representation the strange module.
\subsection{Strange module}\label{Stranger}
Our goal is to describe the action of $\mathfrak{e}$, $\mathfrak{f}$, $\mathfrak{h}$ and $\mathfrak{\psi}$ currents on this strange module. The first obvious remark is that while there is no highest weight vector, nevertheless Cartan currents $\mathfrak{h}(u)$ and $\mathfrak{\psi}(u)$ still can be diagonalized. Let us consider the first component of tensor product $\mathcal{F}_{\boldsymbol{x}}\otimes \mathcal{F}_{\boldsymbol{x}}^{\star}$. We already know \cite{Litvinov:2020zeq} that the eigenbasis is numerated by the collection of Young diagrams $\vec{\boldsymbol{\lambda}}=\{\boldsymbol{\lambda}^{(1)},\dots,\boldsymbol{\lambda}^{(N)}\}$ with the eigenvalues:
\begin{equation}\label{psi-h-eigenvalues}
 h(u)|\vec{\boldsymbol{\lambda}}\rangle=\prod_{\Box\in\vec{\boldsymbol{\lambda}}}\frac{(u-c_{\Box})}{(u-c_{\Box}-\epsilon_{3})}|\vec{\boldsymbol{\lambda}}\rangle,\qquad
	\psi(u)|\vec{\boldsymbol{\lambda}}\rangle=\prod_{\alpha=1}^{3}\prod_{\Box\in\vec{\boldsymbol{\lambda}}}\frac{(u-c_{\Box}-\epsilon_{\alpha})}{(u-c_{\Box}+\epsilon_{\alpha})}
	\prod_{k=1}^{n}\frac{(u-x_{k}+\epsilon_{3})}{(u-x_{k})}|\vec{\boldsymbol{\lambda}}\rangle,
\end{equation}
where by definition  the content of the cell with coordinates $(i,j)$ in Young diagram $\boldsymbol{\lambda}^{(k)}$ is
\begin{equation}
 c_{\Box}=x_{k}-(i-1)\epsilon_{1}-(j-1)\epsilon_{2}.
\end{equation}

Now, both $\mathfrak{h}$ and $\psi$ act by a triangle matrices in the tensor product of two eigenbases. Indeed, according to \eqref{Loper}:
\begin{equation}
    \mathfrak{h}(u)=h(u)\otimes h(-u)+ \sum\limits_{\boldsymbol{\rho} \ne \varnothing} {\mathcal{L}}_{\varnothing,\boldsymbol{\rho}}(u)\otimes \bar{\mathcal{L}}_{\boldsymbol{\rho},\varnothing}(u),
\end{equation}
and hence the eigenbasis of $\mathfrak{h}(u),\psi(u)$ in $\mathcal{F}_{\boldsymbol{x}}\otimes \mathcal{F}^{\star}_{\boldsymbol{x}}$ is enumerated by the collection of $2N$ Young diagrams
\begin{gather}
 \mathfrak{h}(u)\Big|\begin{matrix} \vec{\boldsymbol{\lambda}}\\ \boldsymbol{\vec{\mu}} \end{matrix}\Big\rangle=\prod_{\Box\in\vec{\boldsymbol{\lambda}}}\frac{(u-c_{\Box})}{(u-c_{\Box}-\epsilon_{3})}\prod_{\Box\in\vec{\boldsymbol{\mu}}}\frac{(u-c_{\Box}-\epsilon_{3})}{(u-c_{\Box})}\Big|\begin{matrix} \vec{\boldsymbol{\lambda}}\\ \boldsymbol{\vec{\mu}} \end{matrix}\Big\rangle,\\
	\psi(u)\Big|\begin{matrix} \vec{\boldsymbol{\lambda}}\\ \boldsymbol{\vec{\mu}} \end{matrix}\Big\rangle=\prod_{\alpha=1}^{3}\left(\prod_{\Box\in\vec{\boldsymbol{\lambda}}}\frac{(u-c_{\Box}-\epsilon_{\alpha})}{(u-c_{\Box}+\epsilon_{\alpha})}\prod_{\Box\in\vec{\boldsymbol{\mu}}}\frac{(u-c_{\Box}+\epsilon_{\alpha})}{(u-c_{\Box}-\epsilon_{\alpha})}\right)
	\prod_{k=1}^{n}\frac{(u-x_{k}+\epsilon_{3})}{(u-x_{k})}\frac{(u+x_{k})}{(u+x_{k}+\epsilon_3)}\Big|\begin{matrix} \vec{\boldsymbol{\lambda}}\\ \boldsymbol{\vec{\mu}} \end{matrix}\Big\rangle,
\end{gather}
with the contents
\begin{align}
 &c_{\Box}=x_{k}-(i-1)\epsilon_{1}-(j-1)\epsilon_{2},\quad \text{for Fock modules},\\
 & c_{\Box}=-\epsilon_3-x_{k}+(i-1)\epsilon_{1}+(j-1)\epsilon_{2},\quad \text{for dual Fock modules}.
 \end{align}
Moreover from the $\mathfrak{e},\mathfrak{h}$ commutation relation
\begin{equation}\label{e-application-trial}
 \mathfrak{h}(u)\mathfrak{e}(v)|\Lambda\rangle=\frac{u-v}{u-v-\epsilon_{3}}\mathfrak{e}(v)\mathfrak{h}_{\Lambda}(u)|\Lambda\rangle-\frac{\epsilon_{3}}{u-v-\epsilon_{3}}\mathcal{L}_{\scriptscriptstyle{\varnothing,\Box}}(u)|\Lambda\rangle,
\end{equation}
it follows that $\mathfrak{e}(u)$ acts on the eigenvectors $\Big|\begin{matrix} \vec{\boldsymbol{\lambda}}\\ \boldsymbol{\vec{\mu}} \end{matrix}\Big\rangle$ with the known poles:
\begin{equation}
\begin{aligned}
    \mathfrak{e}(u)\Big|\begin{matrix} \vec{\boldsymbol{\lambda}}\\ \boldsymbol{\vec{\mu}} \end{matrix}\Big\rangle=\sum_{\Box\in\textrm{addable}(\vec{\boldsymbol{\lambda}})}\frac{E\Big(\begin{matrix}\vec{\boldsymbol{\lambda}} &\to &\vec{\boldsymbol{\lambda}}+\Box \\ \vec{\boldsymbol{\mu}}&\to& \vec{\boldsymbol{\mu}} \end{matrix} \Big)}{u-c_{\Box}}
	\Big|\begin{matrix}\vec{\boldsymbol{\lambda}}+\Box\\{\vec{\boldsymbol{\mu}}}\end{matrix}\Big\rangle+\sum_{\Box\in\textrm{removable}(\vec{\boldsymbol{\mu}})}\frac{E\Big(\begin{matrix}\vec{\boldsymbol{\lambda}} &\to &\vec{\boldsymbol{\lambda}}\\ \vec{\boldsymbol{\mu}}&\to& \vec{\boldsymbol{\mu}}-\Box \end{matrix} \Big)}{u-c_{\Box}}
	\Big|\begin{matrix}\vec{\boldsymbol{\lambda}}\\\vec{\boldsymbol{\mu}}-\Box \end{matrix} \Big\rangle, \\
		\mathfrak{f}(u)\Big|\begin{matrix} \vec{\boldsymbol{\lambda}}\\ \boldsymbol{\vec{\mu}} \end{matrix}\Big\rangle=\sum_{\Box\in\textrm{removable}(\vec{\boldsymbol{\lambda}})}\frac{F\Big(\begin{matrix}\vec{\boldsymbol{\lambda}} &\to &\vec{\boldsymbol{\lambda}}-\Box \\ \vec{\boldsymbol{\mu}}&\to& \vec{\boldsymbol{\mu}} \end{matrix} \Big)}{u-c_{\Box}}
	\Big|\begin{matrix}\vec{\boldsymbol{\lambda}}-\Box\\{\vec{\boldsymbol{\mu}}}\end{matrix}\Big\rangle+\sum_{\Box\in\textrm{addable}(\vec{\boldsymbol{\mu}})}\frac{F\Big(\begin{matrix}\vec{\boldsymbol{\lambda}} &\to &\vec{\boldsymbol{\lambda}} \\ \vec{\boldsymbol{\mu}}&\to& \vec{\boldsymbol{\mu}}+\Box \end{matrix} \Big)}{u-c_{\Box}}
		\Big|\begin{matrix}\vec{\boldsymbol{\lambda}}\\\vec{\boldsymbol{\mu}}+\Box \end{matrix} \Big\rangle. 
	\end{aligned}
	\end{equation}
We have a freedom to change the coefficients $F,E$ by re-scaling the eigenvectors, however their product is fixed by the $\mathfrak{ef}$ commutation relation \eqref{ef-relation}:
\begin{equation} \label{EF1}
E\Big(\begin{matrix}\vec{\boldsymbol{\lambda}}-\Box &\to &\vec{\boldsymbol{\lambda}} \\ \vec{\boldsymbol{\mu}}&\to& \vec{\boldsymbol{\mu}} \end{matrix} \Big)F\Big(\begin{matrix}\vec{\boldsymbol{\lambda}} &\to &\vec{\boldsymbol{\lambda}}-\Box\\ \vec{\boldsymbol{\mu}}&\to& \vec{\boldsymbol{\mu}} \end{matrix} \Big)=\textrm{Res}_{u=c_{\Box}}\frac{\Big \langle \begin{matrix} \vec{\boldsymbol{\lambda}} \\ \vec{\boldsymbol{\mu}}\end{matrix}\Big|\psi(u) \Big|\begin{matrix} \vec{\boldsymbol{\lambda}} \\ \vec{\boldsymbol{\mu}}\end{matrix}\Big\rangle}{\Big \langle \begin{matrix} \vec{\boldsymbol{\lambda}} \\ \vec{\boldsymbol{\mu}}\end{matrix}\Big|\begin{matrix} \vec{\boldsymbol{\lambda}} \\ \vec{\boldsymbol{\mu}}\end{matrix}\Big\rangle}
\end{equation}
and
\begin{equation} \label{EF2}
E\Big(\begin{matrix}\vec{\boldsymbol{\lambda}} &\to &\vec{\boldsymbol{\lambda}} \\ \vec{\boldsymbol{\mu}}+\Box &\to& \vec{\boldsymbol{\mu}} \end{matrix} \Big)F\Big(\begin{matrix}\vec{\boldsymbol{\lambda}} &\to &\vec{\boldsymbol{\lambda}}\\ \vec{\boldsymbol{\mu}}&\to& \vec{\boldsymbol{\mu}}+\Box \end{matrix} \Big)=\textrm{Res}_{u=c_{\Box}}\frac{\Big \langle \begin{matrix} \vec{\boldsymbol{\lambda}} \\ \vec{\boldsymbol{\mu}}\end{matrix}\Big|\psi(u) \Big|\begin{matrix} \vec{\boldsymbol{\lambda}} \\ \vec{\boldsymbol{\mu}}\end{matrix}\Big\rangle}{\Big \langle \begin{matrix} \vec{\boldsymbol{\lambda}} \\ \vec{\boldsymbol{\mu}}\end{matrix}\Big|\begin{matrix} \vec{\boldsymbol{\lambda}} \\ \vec{\boldsymbol{\mu}}\end{matrix}\Big\rangle}
\end{equation}

The choice of coefficients $E$ and $F$ consistent with \eqref{EF1}-\eqref{EF2} is equivalent to the choice of normalisation for eigenvectors. It is convenient to use the following one 
\begin{align}\label{E-amplitude-explicit} 
	&E\Big(\begin{matrix}\vec{\boldsymbol{\lambda}} &\to &\vec{\boldsymbol{\lambda}}+\Box \\ \vec{\boldsymbol{\mu}}&\to& \vec{\boldsymbol{\mu}} \end{matrix} \Big)=\frac{\epsilon_{1}\epsilon_{2}}{\epsilon_{3}}\prod_{\Box'\in\vec{\boldsymbol{\lambda}}+\Box}
	S^{-1}(c_{\Box}-c_{\Box'})\prod_{\Box'\in\vec{\boldsymbol{\mu}}}
	S(c_{\Box}-c_{\Box'})\prod_{k=1}^{n}\frac{(c_{\Box}-x_{k}+\epsilon_{3})}{(c_{\Box}-x_{k})}\frac{(c_{\Box}+x_{k})}{(c_{\Box}+x_{k}+\epsilon_3)},\\ \label{F1-amplitude-explicit}
	&F\Big(\begin{matrix}\vec{\boldsymbol{\lambda}} &\to &\vec{\boldsymbol{\lambda}} \\ \vec{\boldsymbol{\mu}}&\to& \vec{\boldsymbol{\mu}}+\Box \end{matrix} \Big)=\prod_{\Box'\in\vec{\boldsymbol{\lambda}}}S(c_{\Box'}-c_{\Box})\prod_{\Box'\in\vec{\boldsymbol{\mu}}+\Box}S^{-1}(c_{\Box'}-c_{\Box}) \prod_{k=1}^{n}\frac{(c_{\Box}-x_{k}+\epsilon_{3})}{(c_{\Box}-x_{k})}\frac{(c_{\Box}+x_{k})}{(c_{\Box}+x_{k}+\epsilon_3)} , \\
	&E\Big(\begin{matrix}\vec{\boldsymbol{\lambda}} &\to &\vec{\boldsymbol{\lambda}} \\ \vec{\boldsymbol{\mu}}&\to& \vec{\boldsymbol{\mu}}-\Box \end{matrix} \Big)=\frac{\epsilon_{1}\epsilon_{2}}{\epsilon_{3}}\prod_{\Box'\in\vec{\boldsymbol{\lambda}}}
	S^{-1}(c_{\Box}-c_{\Box'})\prod_{\Box'\in\vec{\boldsymbol{\mu}-\Box}}
	S(c_{\Box}-c_{\Box'}),\\
	&F\Big(\begin{matrix}\vec{\boldsymbol{\lambda}} &\to &\vec{\boldsymbol{\lambda}} -\Box\\ \vec{\boldsymbol{\mu}}&\to& \vec{\boldsymbol{\mu}} \end{matrix} \Big)=\prod_{\Box'\in\vec{\boldsymbol{\lambda}}-\Box}S(c_{\Box'}-c_{\Box})\prod_{\Box'\in\vec{\boldsymbol{\mu}}}S^{-1}(c_{\Box'}-c_{\Box}), \label{F-amplitude-explicit}
	\end{align}
	with 
	 \begin{equation}\label{S-function-def}
		S(x)=\frac{(x+\epsilon_1)(x+\epsilon_2)}{x(x-\epsilon_3)}.
	 \end{equation}

Let us now define the vector $|\chi\rangle_{\boldsymbol{x}}$ announced in definitions of off-shell Bethe vector \eqref{definition-off-shell-bethe-vector}, \eqref{Bethe-vector-rotated-picture}. The idea is to choose the vector which will maximally simplify the computation of Bethe vector. The most natural definition is:
\begin{equation}\label{chi-definition}
\Big|\begin{matrix} \chi \\ \varnothing\end{matrix}\Big\rangle_{\boldsymbol{x}}=\Big|\begin{matrix} \Box,\dots, \Box \\ \varnothing\end{matrix}\Big\rangle_{\boldsymbol{x}}.
\end{equation}
Alternatively, this vector may be defined (up to proportionality constant) as an eigenvector of $\mathfrak{h}(z)$ with the most natural eigenvalue:
\begin{equation}
\mathfrak{h}(u)\Big|\begin{matrix} \chi \\ \varnothing\end{matrix}\Big\rangle_{\boldsymbol{x}}=\prod\limits_{i=1}^N\frac{u-x_i}{u-x_i-\epsilon_3}\Big|\begin{matrix} \chi \\ \varnothing\end{matrix}\Big\rangle_{\boldsymbol{x}}.
\end{equation}
The main advantage of this choice is that it provides an understandable structure of the off-shell Bethe function \eqref{off-shell-function}, \eqref{off-shell-function-rewritten}. Namely the matrix elements
\begin{equation}
    _{\boldsymbol{x}}\hspace*{-1pt}\langle\mathcal{K^{\alpha}}|
\mathfrak{h}(u_n)\underbrace{\mathfrak{f}(z_1^{(n)})\mathfrak{f}(z_2^{(n)})\dots}_{|\boldsymbol{\lambda}^{(n)}|}\,\,\dots\,\, \mathfrak{h}(u_2)\underbrace{\mathfrak{f}(z_1^{(2)})\mathfrak{f}(z_2^{(2)})\dots}_{|\boldsymbol{\lambda}^{(2)}|}\,\mathfrak{h}(u_1)\underbrace{\mathfrak{f}(z_1^{(1)})\mathfrak{f}(z_2^{(1)})\dots }_{|\boldsymbol{\lambda}^{(1)}|}\Big|\begin{matrix} \chi \\ \varnothing \end{matrix}\Big\rangle_{\boldsymbol{x}}
\end{equation}
involved in \eqref{off-shell-function-rewritten} may have poles only at points $z_i =x_j\ \text{or}\ z_i=-x_j-\epsilon_3$. Then one may compute them explicitly either from formulas \eqref{F1-amplitude-explicit}, \eqref{F-amplitude-explicit}, \eqref{K-amplitude-explicit}, or by analysis of $\mathfrak{hf},\mathfrak{ff}$ commutation relations \eqref{hh-relation}, \eqref{ff-exact-relation} and $\langle\mathcal{K}|\mathfrak{f}$ relation \eqref{Kf}. This logic will be explained in section \eqref{of-shell-computation}.
\subsection{Calculation of K-operator}
Our $K-$operator $\mathcal{K}_{\boldsymbol{x}}$ provides the pairing in the space $\mathcal{F}_{\boldsymbol{x}}\otimes \mathcal{F}_{\boldsymbol{x}}^{\star}$. Our goal for this section is the calculation of the matrix elements:
\begin{equation}
  \langle\vec{\boldsymbol{\mu}}|  \mathcal{K}_{\boldsymbol{x}} |\vec{\boldsymbol{\lambda}}\rangle=\Big\langle \mathcal{K}_{\boldsymbol{x}}\Big|\begin{matrix}\vec{\boldsymbol{\lambda}} \nonumber \\ \vec{\boldsymbol{\mu}} \end{matrix}\Big\rangle
\end{equation}
In order to do so, we use the reflection equation:
\begin{equation}\label{Kmatrix_el}
    K_v \bar{\mathcal{L}}(u) \mathcal{K}_x \mathcal{L}(u)=\mathcal{L}(u) \mathcal{K}_x \bar{\mathcal{L}}(u) K_u 
\end{equation}
Being rewritten in terms of $\mathfrak{L}(u)$, equation \eqref{Kmatrix_el} takes the form:
\begin{equation}
\langle\mathcal{K}| \mathfrak{L}(u)_{\boldsymbol{\lambda},\mathcal{K}\boldsymbol{\mu}}=\langle\mathcal{K}| \bar{\mathfrak{L}}(u)_{\mathcal{K}\boldsymbol{\lambda},\boldsymbol{\mu}}
\end{equation}
Two immediate consequences of these relations are:
\begin{gather}
\langle\mathcal{K}| \mathfrak{h}(u)=\langle\mathcal{K}| \mathfrak{h}(-u), \label{Kh}\\ \label{KL}
\langle\mathcal{K}| \mathfrak{L}_{\Box,\varnothing}(u)=-\kappa (u)\langle\mathcal{K}| \mathfrak{L}_{\Box,\varnothing}(-u),
\end{gather}
where $\mathcal{K}|\Box\rangle=\kappa(u)|\Box\rangle$.
The last equation can be equivalently rewritten in terms of the reflection relation for the $\mathfrak{f}$ current:
\begin{gather} \label{Kf}
\langle \mathcal{K}|\mathfrak{f}(u)=r(u) \langle \mathcal{K}| \mathfrak{f}(-\epsilon_3-u),
\end{gather}
with
\begin{equation}
    r(u)=-\frac{2u+\epsilon_3+\epsilon_3 \kappa (u)}{2u\kappa(u)}.
\end{equation}
This equation immediately follows from \eqref{Kh}, \eqref{KL} after substitution $\mathfrak{L}_{\Box,\varnothing}(u)=\mathfrak{f}(u)\mathfrak{h}(u)$ and the  following chain of relations
\begin{align*}
\langle\mathcal{K}|\mathfrak{f}(u)\mathfrak{h}(u)=-\kappa(u)\langle\mathcal{K}|\mathfrak{f}(-u)\mathfrak{h}(-u)=-\kappa (u) \langle\mathcal{K}|\mathfrak{h}(-u)\mathfrak{f}(-u-\epsilon_3)=\\=-\kappa (u) \langle\mathcal{K}|\mathfrak{h}(u)\mathfrak{f}(-u-\epsilon_3)=-\frac{\epsilon_3 \kappa (u)}{2u+\epsilon_3}\langle\mathcal{K}|\mathfrak{f}(u)\mathfrak{h}(u)-\frac{2u\kappa (u)}{2u+\epsilon_3}\langle\mathcal{K}|\mathfrak{f}(-u-\epsilon_3)h(u)
\end{align*}
Finally we have
\begin{gather}
     \kappa(u)=1 \,, \quad r(u-\epsilon_3/2)=-\frac{u+\epsilon_3/2}{u-\epsilon_3/2} \quad \text{for the \textrm{D} case}\label{rD},\\
     \kappa(u)=\frac{u-\epsilon_i-\epsilon_j/2}{u+\epsilon_i+\epsilon_j/2} \,, \quad r(u-\epsilon_3/2)=-\frac{u+\epsilon_i/2}{u-\epsilon_i/2} \quad \text{for the \textrm{BC} case}, \label{rBC}
\end{gather}
where in the last line $\{i,j\}=\{1,2\}$ corresponds to the \textrm{B} case and $\{i,j\}=\{2,1\}$ corresponds to the \textrm{C} case.

Relations \eqref{Kh} and \eqref{rD}-\eqref{rBC} completely define the matrix elements \eqref{Kmatrix_el}. First of all from \eqref{Kh} it follows that $\mathcal{K}$ acts diagonally in the eigenbasis of $\mathfrak{h}$ i.e. $\vec{\boldsymbol{\lambda}}=\vec{\boldsymbol{\mu}}$. Then one can find
\begin{gather}
    \Big\langle \mathcal{K}_{\boldsymbol{x}}\Big|\begin{matrix}\vec{\boldsymbol{\lambda}} \nonumber \\ \vec{\boldsymbol{\lambda}} \end{matrix}\Big\rangle =F^{-1}\Big(\begin{matrix}\vec{\boldsymbol{\lambda}}  & \to & \vec{\boldsymbol{\lambda}}\\ \vec{\boldsymbol{\lambda}}-\Box & \to & \vec{\boldsymbol{\lambda}}\end{matrix} \Big)\textrm{res}_{z=c_{\Box}}\Big\langle \mathcal{K}_{\boldsymbol{x}}\Big| f(-\epsilon_3-z) \Big|\begin{matrix}\vec{\boldsymbol{\lambda}}  \\ \vec{\boldsymbol{\lambda}}-\Box \end{matrix}\Big\rangle=\\=r^{-1}(c_{\Box})F^{-1}\Big(\begin{matrix}\vec{\boldsymbol{\lambda}}  & \to & \vec{\boldsymbol{\lambda}}\\ \vec{\boldsymbol{\lambda}}-\Box & \to & \vec{\boldsymbol{\lambda}}\end{matrix} \Big)\textrm{res}_{z=c_{\Box}} \Big\langle\mathcal{K}_{\boldsymbol{x}}\Big| f(z) \Big|\begin{matrix}\vec{\boldsymbol{\lambda}} \\ \label{K-amplitude-explicit} \vec{\boldsymbol{\lambda}}-\Box \end{matrix}\Big\rangle=\\=r^{-1}(c_{\Box})F^{-1}\Big(\begin{matrix}\vec{\boldsymbol{\lambda}}  & \to & \vec{\boldsymbol{\lambda}}\\ \vec{\boldsymbol{\lambda}}-\Box & \to & \vec{\boldsymbol{\lambda}}\end{matrix} \Big)F\Big(\begin{matrix}\vec{\boldsymbol{\lambda}}  & \to & \vec{\boldsymbol{\lambda}}-\Box\\ \nonumber \vec{\boldsymbol{\lambda}}-\Box & \to & \vec{\boldsymbol{\lambda}}-\Box \end{matrix} \Big) \Big\langle\mathcal{K}_{\boldsymbol{x}} \Big|\begin{matrix}\vec{\boldsymbol{\lambda}}-\Box \\ \vec{\boldsymbol{\lambda}}-\Box \end{matrix}\Big\rangle
\end{gather}
\subsection{Off-shell Bethe function, diagonalization of KZ integral.}\label{of-shell-computation}
Motivated by the formulas \eqref{rD} and \eqref{rBC}, it is convenient to shift $x$ variables: $x\to x-\frac{\epsilon_3}{2}$, as well as redefine the operators $\mathfrak{f}$: $\mathfrak{f}(z)\to\mathfrak{f}(z-\frac{\epsilon_3}{2})$.

Let us consider the following Bethe vectors:
\begin{align}
&|B^{\alpha,\boldsymbol{u}}(\boldsymbol{x})\rangle=_{\boldsymbol{x}}\! \! \langle\varnothing|\bar{\mathcal{L}}_1\dots\bar{\mathcal{L}}_n \mathcal{K}^{\alpha}_{\boldsymbol{x}} \mathcal{L}_n \dots \mathcal{L}_1 |\varnothing\rangle_{\boldsymbol{u}}  |\chi\rangle_{\boldsymbol{x}} \equiv _{\boldsymbol{x}}\! \! \langle \mathcal{K}^{\alpha}|\mathfrak{L}_n\dots \mathfrak{L}_1 |\varnothing\rangle_{\boldsymbol{u}}\Big|\begin{matrix} \chi \\ \varnothing \end{matrix}\Big\rangle_{\boldsymbol{x}}  \\
&|\bar{B}^{\beta,\boldsymbol{u}}(\boldsymbol{x})\rangle=_{\boldsymbol{x}}\! \! \langle \varnothing| \mathcal{L}_n \dots \mathcal{L}_1 \mathcal{K}^{\beta}_{\boldsymbol{x}} \bar{\mathcal{L}}_1\dots\bar{\mathcal{L}}_n |\varnothing\rangle_{\boldsymbol{u}}|\bar{\chi}\rangle_{\boldsymbol{x}} 
\equiv _{\boldsymbol{x}}\! \! \langle \mathcal{K}^{\beta}|\bar{\mathfrak{L}}_1\dots \bar{\mathfrak{L}}_n |\varnothing\rangle_{\boldsymbol{u}}\Big|\begin{matrix} \bar{\chi} \\ \varnothing \end{matrix}\Big\rangle_{\boldsymbol{x}} ,
\end{align}
where $\alpha=1,2,3$ labels possible $K-$operators.

It is also useful to introduce their matrix elements the so called off-shell Bethe functions:
\begin{align}\label{off-shell-function}
\omega_{\alpha,\vec{\boldsymbol{\lambda}}}(\boldsymbol{x}|\boldsymbol{u})\overset{\text{def}}{=}\langle \vec{\boldsymbol{\lambda}}|B^{\alpha,\boldsymbol{u}}(\boldsymbol{x})\rangle=_{\boldsymbol{x}}\hspace*{-4pt}\langle \mathcal{K}^{\alpha}_x|
	\mathfrak{L}_{\boldsymbol{\lambda}^{(1)},\varnothing}(u_{1})\dots\mathfrak{L}_{\boldsymbol{\lambda}^{(n)},\varnothing}(u_{n})\Big|\begin{matrix} \chi \\ \varnothing \end{matrix}\Big\rangle _{\boldsymbol{x}}, \\
\bar{\omega}_{\beta,\vec{\boldsymbol{\lambda}}}(\boldsymbol{x}|\boldsymbol{u})\overset{\text{def}}{=}\langle \vec{\boldsymbol{\lambda}}|\bar{B}^{\beta,\boldsymbol{u}}(\boldsymbol{x})\rangle=_{\boldsymbol{x}}\hspace*{-4pt}\langle \mathcal{K}^{\beta}_x|
	\bar{\mathfrak{L}}_{\boldsymbol{\lambda}^{(n)},\varnothing}(u_{n})\dots\bar{\mathfrak{L}}_{\boldsymbol{\lambda}^{(1)},\varnothing}(u_{1})\Big|\begin{matrix} \bar{\chi} \\ \varnothing \end{matrix}\Big\rangle_{\boldsymbol{x}}.
\end{align}
The off-shell Bethe vectors and hence the off-shell functions have  nice intertwining relations with $R$-matrix and $K$-operators:
\begin{align}
\mathcal{R}_{i,i+1}|B^{\alpha,\boldsymbol{u}}(\boldsymbol{x})\rangle=P_{i,i+1}|B^{\alpha,\boldsymbol{u}}(\boldsymbol{x})\rangle \label{IntertwiningRels1},\\
\mathcal{R}_{i,i+1}|\bar{B}^{\beta,\boldsymbol{u}}(\boldsymbol{x})\rangle=P_{i,i+1}|\bar{B}^{\beta,\boldsymbol{u}}(\boldsymbol{x})\rangle,\\
\mathcal{K}^{\alpha}_{n}|B^{\alpha,\boldsymbol{u}}(\boldsymbol{x})\rangle=D_n|B^{\alpha,\boldsymbol{u}}(\boldsymbol{x})\rangle,\\
\mathcal{K}^{\beta}_{1}D_1|\bar{B}^{\beta,\boldsymbol{u}}(\boldsymbol{x})\rangle=|\bar{B}^{\beta,\boldsymbol{u}}(\boldsymbol{x})\rangle. \label{IntertwiningRels4}
\end{align}
It implies in particular, the simple action of  \textrm{KZ}  operators:
\begin{align} \label{Intertwinting-of-BetheVector}
\mathcal{T}^+_i |B^{\alpha,\boldsymbol{u}}(\boldsymbol{x})\rangle=D_i|B^{\alpha,\boldsymbol{u}}(\boldsymbol{x})\rangle, \\
\label{Intertwinting-of-BarBetheVector}
\mathcal{T}^-_iD_i|\bar{B}^{\beta,\boldsymbol{u}}(\boldsymbol{x})\rangle=|\bar{B}^{\beta,\boldsymbol{u}}(\boldsymbol{x})\rangle.
\end{align}

Our goal for this section is to prove that under the Bethe ansatz equations \eqref{BAE} two Bethe vectors are proportional to each other:
\begin{equation}\label{B=barB}
|B^{\alpha,\boldsymbol{u}}(\boldsymbol{x})\rangle \overset{\text{BAE}(\boldsymbol{x})=1}{=}|\bar{B}^{\alpha,\boldsymbol{u}}(\boldsymbol{x})\rangle \prod\limits_{i,a}\frac{u_i-x_a-\frac{\epsilon_3}{2}}{u_i-x_a+\frac{\epsilon_3}{2}} \ c({\boldsymbol{x}}),
\end{equation}
for some $c(\boldsymbol{x})$. Combining this equation and relations \eqref{Intertwinting-of-BetheVector}-\eqref{Intertwinting-of-BarBetheVector} one can immediately conclude that on-shell Bethe vector is indeed an eigenvector of \textrm{KZ} operator \eqref{KZeigenvalue}.

Let us proceed to the proof. It is enough to check \eqref{B=barB} for any matrix element, or, which is the same, to establish similar relation for off-shell functions \eqref{w=barw}. In order to compute the later, let us remind \eqref{LAsSh} that $\mathfrak{L}_{\boldsymbol{\lambda},\varnothing}$ is generated by the $\mathfrak{h}(u)$ and $\mathfrak{f}(z)$ currents:
\begin{equation}\label{matrix-element-Lax-integral-representaion}
	\mathfrak{L}_{\boldsymbol{\lambda},\varnothing}(u)=\frac{1}{(2\pi i)^{|\boldsymbol{\lambda}|}}\oint_{\mathcal{C}_1}\dots\oint_\mathcal{C_{|\boldsymbol{\lambda}|}} F_{\boldsymbol{\lambda}}(\boldsymbol{z}-\frac{\epsilon_3}{2}|u)\,\mathfrak{h}(u)\mathfrak{f}(z_{|\boldsymbol{\lambda}|})\dots \mathfrak{f}(z_1)dz_1\dots dz_{|\boldsymbol{\lambda}|},
\end{equation} 
where each contour $\mathcal{C}_k$ goes clockwise around $\infty$ and $u-\epsilon_{3}$, so that it doesn't pick the poles of function $F_{\boldsymbol{\vec{\lambda}}}$. Using \eqref{matrix-element-Lax-integral-representaion} the weight function \eqref{off-shell-function} can be rewritten as
\begin{multline}\label{off-shell-function-rewritten}
\omega_{\alpha,\vec{\boldsymbol{\lambda}}}(\boldsymbol{x}|\boldsymbol{u})=\frac{1}{(2\pi i)^{N}}\times\\\times\oint
F_{\vec{\boldsymbol{\lambda}}}(\vec{\boldsymbol{z}}-\frac{\epsilon_3}{2}|\boldsymbol{u})\,\,
_{\boldsymbol{x}}\hspace*{-1pt}\langle\mathcal{K^{\alpha}}|
\mathfrak{h}(u_n)\underbrace{\mathfrak{f}(z_1^{(n)})\mathfrak{f}(z_2^{(n)})\dots}_{|\boldsymbol{\lambda}^{(n)}|}\,\,\dots\,\, \mathfrak{h}(u_2)\underbrace{\mathfrak{f}(z_1^{(2)})\mathfrak{f}(z_2^{(2)})\dots}_{|\boldsymbol{\lambda}^{(2)}|}\,\mathfrak{h}(u_1)\underbrace{\mathfrak{f}(z_1^{(1)})\mathfrak{f}(z_2^{(1)})\dots }_{|\boldsymbol{\lambda}^{(1)}|}\Big|\begin{matrix} \chi \\ \varnothing \end{matrix}\Big\rangle_{\boldsymbol{x}}\,d\vec{\boldsymbol{z}},
\end{multline}
where
\begin{equation}\label{F-lambda-vec}
F_{\vec{\boldsymbol{\lambda}}}(\vec{\boldsymbol{z}}|\boldsymbol{u})=\prod_{k=1}^{n}F_{\boldsymbol{\lambda}^{(k)}}\left(z_1^{(k)},\dots,z^{(k)}_{|\boldsymbol{\lambda}^{(k)}|}\bigr|u_k\right).
\end{equation}

The contour integral \eqref{off-shell-function-rewritten} can be computed by residues. In order to do so let us analyse possible poles in $z$ variables. As we already explained any $\mathfrak{f}(z)$ operator either removes one box from the upper Young diagram $\chi$ or add a box to the lover Young diagram with a pole equal to the content of the corresponding cell. We also proved that the matrix element $\Big\langle \mathcal{K}\Big|\begin{matrix}\boldsymbol{\vec{\lambda}}\\\boldsymbol{\vec{\mu}}\end{matrix}\Big\rangle$ is nonzero only if $\boldsymbol{\vec{\lambda}}=\boldsymbol{\vec{\mu}}$. Thus we conclude that the only possible poles of the contour integral \eqref{off-shell-function-rewritten} are
\begin{equation}
z_i=x_{\sigma(i)} \quad \text{or} \quad z_i=-x_{\sigma(i)},    
\end{equation}
where $\sigma$ is some permutation. It is convenient to consider the group spanned by the elements $s$, which is generated by permutations of all indices $S_N$ and $\mathbb{Z}_2$ reflection of each index $i\to \bar{i}$ with the convention $x_{\bar{i}}=-x_i$. The weight function itself is given by the sum over residues as:
\begin{equation}
    \omega_{\alpha,\vec{\boldsymbol{\lambda}}}(\boldsymbol{x}|\boldsymbol{u})=\sum_{s} I^{\alpha}_{s(1,\dots,N)} \ .
\end{equation}
We are going to prove the proportionality of two weight functions under the Bethe equations:
\begin{equation}\label{w=barw}
     \omega_{\alpha,\vec{\boldsymbol{\lambda}}}(\boldsymbol{x}|\boldsymbol{u})\overset{\text{BAE}(\boldsymbol{x})=1}{=}c(\boldsymbol{x}) \bar{\omega}_{\beta,\vec{\boldsymbol{\lambda}}}(\boldsymbol{x}|\boldsymbol{u}).
\end{equation}
Actually, we will prove a stronger statement of proportionality of the corresponding residues:
\begin{equation}\label{I=I}
    I^{\alpha}_{s(1,\dots,N)}\overset{\text{BAE}(\boldsymbol{x})=1}{=}c(\boldsymbol{x}) \bar{I}^{\beta}_{s(1,\dots,N)}.
\end{equation}
Using the fact that the integral \eqref{matrix-element-Lax-integral-representaion} defined to avoid the poles of $F_{\boldsymbol{\vec{\lambda}}}$, we have explicitly:
\begin{multline}\label{I-explicit}
I^{\alpha}_{s(1,\dots N)}(\boldsymbol{x})=F_{\vec{\boldsymbol{\lambda}}}(s(\boldsymbol{x})-\frac{\epsilon_3}{2}|\boldsymbol{u})
\,\text{Res}_{z_i=x_{s(i)}}\ _{\boldsymbol{x}}\hspace*{-1pt}\langle\mathcal{K^{\alpha}}|
\mathfrak{h}(u_n)\underbrace{\mathfrak{f}(z_1^{(n)})\mathfrak{f}(z_2^{(n)})\dots}_{|\boldsymbol{\lambda}^{(n)}|}\,\,\dots\\\dots\,\, \mathfrak{h}(u_2)\underbrace{\mathfrak{f}(z_1^{(2)})\mathfrak{f}(z_2^{(2)})\dots}_{|\boldsymbol{\lambda}^{(2)}|}\,\mathfrak{h}(u_1)\underbrace{\mathfrak{f}(z_1^{(1)})\mathfrak{f}(z_2^{(1)})\dots }_{|\boldsymbol{\lambda}^{(1)}|}\Big|\begin{matrix} \chi \\ \varnothing \end{matrix}\Big\rangle_{\boldsymbol{x}}
\end{multline}
and
\begin{multline}\label{barI-explicit}
\bar{I}^{\beta}_{s(1,\dots N)}(\boldsymbol{x})=F_{\vec{\boldsymbol{\lambda}}}(s(\boldsymbol{x})-\frac{\epsilon_3}{2}|\boldsymbol{u})\,\text{Res}_{z_i=x_{s(i)}}\cdot\\ 
_{\boldsymbol{x}}\hspace*{-1pt}\langle\mathcal{K^{\beta}}|
\underbrace{\mathfrak{f}(-z_1^{(1)})\mathfrak{f}(-z_2^{(1)})\dots}_{|\boldsymbol{\lambda}^{(1)}|}\mathfrak{h}(-u_1)\underbrace{\mathfrak{f}(-z_1^{(2)})\mathfrak{f}(-z_2^{(2)})\dots}_{|\boldsymbol{\lambda}^{(2)}|}\mathfrak{h}\,(-u_2)\,\,\dots\,\,\underbrace{\mathfrak{f}(-z_1^{(n)})\mathfrak{f}(-z_2^{(n)})\dots }_{|\boldsymbol{\lambda}^{(n)}|}\mathfrak{h}(-u_n)\Big|\begin{matrix} \chi \\ \varnothing \end{matrix}\Big\rangle_{\boldsymbol{x}}.
\end{multline}
Taking into account formulas \eqref{E-amplitude-explicit}-\eqref{F-amplitude-explicit} and \eqref{K-amplitude-explicit}, it is straightforward to compute matrix elements and check \eqref{I=I}
with
\begin{equation}\label{propconst}
    c(\boldsymbol{x})=\prod\limits_{i<j}S(x_i+x_j)\prod_i r^{\beta}(-x_i).
\end{equation}

One can also provide a simpler proof without reference to the explicit formulas for matrix elements, but using the $\mathfrak{fh},\mathfrak{ff}$ commutation relation \eqref{hh-relation},\eqref{ff-exact-relation} and $\langle\mathcal{K}|\mathfrak{f}$ relation \eqref{Kf}. The direct consequence of this relations is the formula for the matrix elements:
\begin{equation}\label{hf-matrix}
\text{Res}_{z_i=x_{i}}\ _{\boldsymbol{x}}  \langle\mathcal{K}^{\alpha}|\dots \mathfrak{f}(z_{i+1})\mathfrak{f}(z_i)\dots \Big|\begin{matrix} \chi \\ \varnothing \end{matrix}\Big\rangle_{\boldsymbol{x}}=G^{-1}(x_{i+1}-x_i)\ \text{Res}_{z_i=x_{i}}\ _x  \langle\mathcal{K}^{\alpha}|\dots \mathfrak{f}(z_{i})\mathfrak{f}(z_{i+1})\dots \Big|\begin{matrix}\chi \\ \varnothing \end{matrix}\Big\rangle_{\boldsymbol{x}},
\end{equation}
\begin{equation}\label{ff-matrix}
\text{Res}_{z_i=x_{i}}\ _{\boldsymbol{x}}  \langle\mathcal{K}^{\alpha}|\dots \mathfrak{f}(z_{i})\mathfrak{h}(u)\dots \Big|\begin{matrix}\chi \\ \varnothing \end{matrix}\Big\rangle_{\boldsymbol{x}}=\frac{u-x_i+\frac{\epsilon_3}{2}}{u-x_i-\frac{\epsilon_3}{2}}\, \text{Res}_{z_i=x_{i}}\ _{\boldsymbol{x}}  \langle\mathcal{K}^{\alpha}|\dots \mathfrak{h}(u)\mathfrak{f}(z_{i})\dots \Big|\begin{matrix} \chi \\ \varnothing \end{matrix}\Big\rangle_{\boldsymbol{x}},
\end{equation}
\begin{align}\label{Kf-matrix}
    \text{Res}_{z_i=x_{i}}\ _{\boldsymbol{x}}  \langle\mathcal{K}^{\alpha}|\mathfrak{f}(-z_N)\dots \Big|\begin{matrix}\chi \\ \varnothing \end{matrix}\Big\rangle_{\boldsymbol{x}}=r(-x_N)\  \text{Res}_{z_i=x_{i}}\ _{\boldsymbol{x}}  \langle\mathcal{K}^{\alpha}|\mathfrak{f}(z_N)\dots \Big|\begin{matrix} \chi \\ \varnothing \end{matrix}\Big\rangle_{\boldsymbol{x}}.
\end{align}
Note that in the first two relations we additionally used the fact that Bethe roots $x_i$ are not in resonance with each other as well as with evaluation parameters $u_k$. It allow us to omit local (blue) terms in \eqref{hf-relation}, \eqref{ff-exact-relation}. These three relations are completely define the residues up to a constant. 

We immediately observe that both matrix elements \eqref{I-explicit},\eqref{barI-explicit} share the same transformation properties under the permutation of $x_i$ variables. At the first glance the transformation under reflection of $x_i$ variables is different. Indeed the reflection of $x_N$ in the first matrix element \eqref{I-explicit} produce a simple factor $r^{\alpha}(x_N)$, while in the opposite matrix element \eqref{barI-explicit} we have to move corresponding operator $\mathfrak{f}(-z_N)$ to the left boundary and back which produce the product of many terms:
\begin{equation}
   r^{\beta}(-x_N)\prod\limits_{k}\frac{x_N^2-(u_k-\frac{\epsilon_3}{2})^2}{x_N^2-(u_k+\frac{\epsilon_3}{2})^2} \prod\limits_{j\ne N}G^{-1}(x_N-x_j)G^{-1}(x_N+x_j).
\end{equation}
Two factors coincide under the Bethe equations \eqref{BAE}. This proves the proportionality of corresponding residues \eqref{I-explicit},\eqref{barI-explicit}. The proportionality constant \eqref{propconst} may be computed along the same lines.

It may be useful to note that the identities \eqref{hf-matrix}-\eqref{Kf-matrix} may be summarized in the following rules for computation of the residues:
\begin{equation}
I^{\alpha}_{1,\dots,N}=F_{\vec{\boldsymbol{\lambda}}}(\boldsymbol{x}+\frac{\epsilon_3}{2}|\boldsymbol{u})\prod\limits_k\prod\limits_{i=|\lambda_k|+1}^N\frac{u_k-x_i+\frac{\epsilon_3}{2}}{u_k-x_i-\frac{\epsilon_3}{2}}\prod\limits_{i<j}S(x_i-x_j),
\end{equation}
\begin{equation}
I^{\alpha}_{\dots,i+1,i,\dots}(\boldsymbol{x})=I^{\alpha}_{\dots,i,i+1,\dots}(P_{i,i+1}\boldsymbol{x}),
\end{equation}
\begin{equation}
I^{\alpha}_{\dots,\bar{N}}(\boldsymbol{x})=I^{\alpha}_{\dots,N}(\boldsymbol{x})\Big|_{x_N\to -x_N}r^{\alpha}(x_N).
\end{equation}

Finally, let us compute the action of \textrm{KZ} integral of motion $\mathcal{I}^{\textrm{KZ}}_i=\mathcal{T}^-_i\mathcal{T}^+_i$ on Bethe vector:
\begin{equation}
 \mathcal{T}^-_i \mathcal{T}^+_i|B^{\alpha,\boldsymbol{u}}(\boldsymbol{x})\rangle=\mathcal{T}^-_i D_i|B^{\alpha,\boldsymbol{u}}(\boldsymbol{x})\rangle\overset{\text{BAE}(\boldsymbol{x})=1}{=}\mathcal{T}^-_iD_i|\bar{B}^{\alpha,\boldsymbol{u}}(\boldsymbol{x})\rangle \frac{u_i+x_a+\frac{\epsilon_3}{2}}{u_i+x_a-\frac{\epsilon_3}{2}}\prod\limits_{j\ne i,a}\frac{u_i-x_a-\frac{\epsilon_3}{2}}{u_i-x_a+\frac{\epsilon_3}{2}} \  c({\boldsymbol{x}})
\end{equation}
\begin{gather}
    \mathcal{T}^-_iD_i|\bar{B}^{\alpha,\boldsymbol{u}}(\boldsymbol{x})\rangle =|\bar{B}^{\alpha,\boldsymbol{u}}(\boldsymbol{x})\rangle\overset{\text{BAE}(\boldsymbol{x})=1}{=} |B^{\alpha,\boldsymbol{u}}(\boldsymbol{x})\rangle\frac{u_i-x_a+\frac{\epsilon_3}{2}}{u_i-x_a-\frac{\epsilon_3}{2}} \prod\limits_{j\ne i,a}\frac{u_j-x_a+\frac{\epsilon_3}{2}}{u_j-x_a-\frac{\epsilon_3}{2}} \ c^{-1}({\boldsymbol{x}}),
\end{gather}
which finally proves
\begin{equation}\label{KZeigenvalue}
\mathcal{I}^{\textrm{KZ}}_i|B^{\alpha,\boldsymbol{u}}(\boldsymbol{x})\rangle\overset{\text{BAE}(\boldsymbol{x})=1}{=}|B^{\alpha,\boldsymbol{u}}(\boldsymbol{x})\rangle\frac{(u_i+\frac{\epsilon_3}{2})^2-x_a^2}{(u_i-\frac{\epsilon_3}{2})^2-x_a^2}.
\end{equation}
\section{Concluding remarks}\label{concl}
In this paper we discovered Bethe ansatz equations for the spectrum of Integrals of Motion in CFT with the $W-$symmetry of $\mathrm{BCD}$ type. There many open questions, which we list in random order.
\paragraph{T-operator.}
We have avoided the construction of the boundary transfer matrix similar to Sklyanin \cite{Sklyanin:1988yz}. The reason is that this object is not well defined for $\mathrm{Y}(\widehat{\mathfrak{gl}}(1))$. Its construction requires the corresponding $R$-matrix to satisfy the property known as crossing unitarity. One can easily show that MO $R$-matrix satisfies two basic properties of
\begin{align}
     &\text{Unitarity}:&& R[\partial\varphi_i-\partial\varphi_j]R[\partial\varphi_j-\partial\varphi_i]=1,\label{R-matrix-properties-initarity}\\
     &\text{T-symmetry}:&& R^t[\partial\varphi_i-\partial\varphi_j]=R[\partial\varphi_i-\partial\varphi_j],\label{R-matrix-properties-T-symmetry}
\end{align}
which follow immediately from the defining relations \eqref{RLL-Miura}. Both \eqref{R-matrix-properties-initarity} and \eqref{R-matrix-properties-T-symmetry} hold level by level and can be easily verified by explicit calculations for lower levels. However, the crossing unitarity property
\begin{equation}\label{R-matrix-properties-Crossing-unitariry}
  R^{t_i}[\partial\varphi_i-\partial\varphi_j]R^{t_i}[\partial\varphi_j-\partial\varphi_i]=1.
\end{equation}
is more subtle, as it mixes different levels and involves infinite sums of matrix elements. It is questionable if one can make it any sense. Even if we believe that \eqref{R-matrix-properties-Crossing-unitariry} holds and try to make a step further, we may conjecture the  following formula for the generating function of integrals of motion (for the $\mathrm{D}$ case)
\begin{equation}
\mathcal{T}(u)=\textrm{Tr}\Big|_0 \left(\mathcal{R}_{\bar{0},1}\dots \mathcal{R}_{\bar{0},n}\mathcal{R}_{0,n}\dots \mathcal{R}_{0,1}\right)
\end{equation}
This formula  requires more accurate definition as it involves divergent summation over infinite dimensional Fock space. In the $\mathrm{A}$ case this divergence has been regularised by an introduction of the twist parameter which preserved the integrability. It is unclear whether such a twist can be introduced in the present case as well. This certainly remains as open interesting question. 
\paragraph{Eigenvalues of local Integrals of Motion.}
Our construction is specially adapted to diagonalization of \textrm{KZ} integral. Diagonalization of local IM's is a separate issue. We note that in the $\mathrm{A}$ case \cite{Litvinov:2020zeq} we provided explicit construction for diagonalization of the simplest non-trivial local IM $\mathbf{I}_2$. In principle, it can be easily generalized for $\mathbf{I}_s$ with $s>2$. 

In the present case, we conjecture the following formula for eigenvalues of the integral $\mathbf{I}_3=\frac{1}{2\pi}\int G_4(x)dx$  corresponding to the local density $G_4$ given by \eqref{G4-density}. Namely, on level $N$ one has an eigenvalue
\begin{equation}\label{I3-eigenvalues}
  \mathbf{I}_3^{\textrm{vac}}+\left(4 N-4\sum_{k=1}^n\frac{u_k^2}{\epsilon_1\epsilon_2}+\frac{\epsilon_1^2+\epsilon_2^2}{3\epsilon_1\epsilon_2}\Big(2n-\frac{\epsilon_{\alpha}+\epsilon_{\beta}}{\epsilon_3}\Big)\right)N+\frac{4}{\epsilon_1\epsilon_2}\left(2n-\frac{\epsilon_{\alpha}+\epsilon_{\beta}}{\epsilon_3}\right)\sum_{k=1}^N x_k^2,
\end{equation} 
where $x_k$'s satisfy Bethe ansatz equations
\begin{equation}\label{BetheAnsatz}
r^{(\alpha)}(x_i)r^{(\beta)}(x_i)A(x_i)A^{-1}(-x_i)\prod_{j\neq i}G(x_i-x_j)G^{-1}(-x_i-x_j)=1,
\end{equation}
with
\begin{equation}
  r^{(\alpha)}(x)=-\frac{x-\frac{\epsilon_{\alpha}}{2}}{x+\frac{\epsilon_{\alpha}}{2}},\qquad
  A(x)=\prod_{k=1}^n\frac{x-u_k+\frac{\epsilon_3}{2}}{x-u_k-\frac{\epsilon_3}{2}}.
\end{equation}
We have confirmed \eqref{I3-eigenvalues} by explicit diagonalization on lower levels and it is interesting to find a proof.
\paragraph{Bullough-Dodd model}
Integrable systems studied in this paper are already non-trivial for $n=1$. Let us consider $\widehat{\textrm{BC}}_1$ system, which is known also as Bullough-Dodd model, or Zhiber-Shabat model. This is the theory of one bosonic field $\varphi$ with the  action
\begin{equation}
    S=\int\left(\frac{1}{8\pi}(\partial_a\varphi)^2+\Lambda\Big(e^{2b\varphi}+e^{-b\varphi}\Big)\right)d^2z.
\end{equation}
According to Zamolodchikov \cite{Zamolodchikov:1989zs}, this theory can be interpreted as $\Phi_{1,2}$ integrable perturbation of CFT (or equivalently as $\Phi_{1,5}$ perturbation). From the general formula \eqref{G4-density} we see that $\mathbf{I}_3$ identically vanishes for BD model. It implies the following identity for the Bethe roots
\begin{equation}
    \sum_{k=1}^Nx_k^2=\frac{1}{12}\left(4u^2-4N-Q\right).
\end{equation}
The first non-trivial integral is $\mathbf{I}_5$ which has the form
\begin{multline}\label{I5-BD-Integral}
    \mathbf{I}_5=\frac{1}{2\pi}\int\Bigg[(\partial\varphi)^6-\frac{5}{4}(\partial\varphi)^4-\frac{5}{2}(b-b^{-1})(2Q^2+1)(\partial^2\varphi)^3+\\+5(3Q^2+1)\left((\partial^2\varphi)^2(\partial\varphi)^2-\frac{1}{12}(\partial^2\varphi)^2\right)+\left(3Q^4+\frac{17Q^2}{4}+\frac{8}{3}\right)(\partial^3\varphi)^2\Bigg]dx.
\end{multline}
Here all densities are Wick ordered. We note that our integral \eqref{I5-BD-Integral} differs from the analytically regularized integral by addition of $\mathbf{I}_1$ and a constant. Bethe Ansatz equations follows the general rules \eqref{BetheAnsatz} with $\alpha=1$, $\beta=2$ and $n=1$. We found that the eigenvalues of $\mathbf{I}_5-\mathbf{I}^{\text{vac}}_5$ are given by
\begin{equation}
  N\left(\frac{63Q^4}{8}+\Big(45N-\frac{63}{2}\Big)Q^2+80N^2-95N+27\right)-5N(9Q^2+24N-19)u^2+60Nu^4-270\sum\limits_{k=1}^{N} x_k^4.
\end{equation}
\paragraph{Colored Fock spaces and more general integrable systems.}
One may wonder that despite affine Yangian commutation relations \eqref{Yangian-relation-main} are symmetric with respect to permutations of all $\epsilon_{\alpha}$, Bethe Ansatz equations \eqref{BAE} are not symmetric in all $\epsilon_{\alpha}$ because of the source term 
\begin{equation}
A(x)=\prod\limits_{k=1}^n\frac{x-u_k+\frac{\epsilon_3}{2}}{x-u_k-\frac{\epsilon_3}{2}}.
\end{equation}
In fact, there exist three types of Fock modules $\mathcal{F}^{\alpha}$ (see \cite{feigin2013representations,2015arXiv151208779B,Litvinov:2016mgi}), introducing them into a game provides us with  more general integrable systems. In fact, we associate an integrable system to the chain of colored Fock spaces with two colored boundaries $\beta_L\Big|\mathcal{F}_1^{\alpha_1}\otimes \mathcal{F}_2^{\alpha_2} \dots \otimes \mathcal{F}_n^{\alpha_n}\Big|\beta_R$ \ , \  $\alpha_i,\beta_{L,R}=1,2,3$.  We present the details in Appendix \ref{3Fock}, here we just mention a particular interesting model given as: $1\Big|\mathcal{F}_1^{1}\otimes \mathcal{F}_2^{3} \dots \otimes \mathcal{F}_{2n-1}^{1}\otimes\mathcal{F}_{2n}^{3}\Big|3$. This model provides a UV limit for the (dual of) $O(2n+1)$ sigma model considered in \cite{Litvinov:2018bou}. Similarly \\  $3\Big|\mathcal{F}_1^{3}\otimes \mathcal{F}_2^{1} \dots \otimes \mathcal{F}_{2n+1}^{3}\Big|3 $ gives the UV limit of $O(2n)$ sigma model.
\paragraph{$K$-matrices.}
We have mentioned in the main text that there are only three solutions of Sklyanin reflection  equation \eqref{reflection-equation}, which commute with the level. In such a case one can always set the vacuum eigenvalue of $\mathcal{K}$ operator to $1$. Then, if we  denote
\begin{equation}
  \mathcal{K}a_{-1}|u\rangle=f(u)a_{-1}|u\rangle,
\end{equation}
the reflection relation \eqref{reflection-equation} on level $1$ is equivalent to the functional relation
\begin{equation}
    (u+v)(f(u)-f(v))=(u-v)(f(u)f(v)-1)\implies f(u)=\frac{\xi+u}{\xi-u},
\end{equation}
where $\xi$ is an arbitrary parameter. The reflection relation on level $2$ is more restrictive. It is not just fixes the matrix of the $K$-operator on level $2$, but also demands that the parameter $\xi$ takes one of three values
\begin{equation}
    \xi=0,\quad\xi=-\left(b+\frac{1}{2b}\right)\quad\text{or}\quad\xi=-\left(\frac{1}{b}+\frac{b}{2}\right),
\end{equation}
corresponding to three solutions $\mathcal{K}^{1,2,3}$. In principle, one might go to higher levels and check that there are only three solutions. It would be interesting to prove this statement in general.
\section*{Acknowledgements}
A.L. acknowledges the support of Basis Foundation. I.V. has been supported in part by Young Russian Mathematics award.
\Appendix
\section{Restoring the symmetry between \texorpdfstring{$\epsilon_{\alpha}$}{epsilon}} \label{3Fock}
One may note that affine Yangian commutation relations \eqref{Yangian-relation-main} are symmetric with respect to permutations of all $\epsilon_{\alpha}$. Nevertheless Bethe Ansatz equations \eqref{BAE} are not symmetric in all $\epsilon_{\alpha}$, because of the source term $A(x)=\prod\limits_{k=1}^n\frac{x-u_k+\frac{\epsilon_3}{2}}{x-u_k-\frac{\epsilon_3}{2}}$. We are now in a position to restore the symmetry, which will help us to build more general integrable systems. The resolution of the paradox is the following: there actually exists three types of Fock modules $\mathcal{F}^{\alpha}_x$. In order to describe Integrable systems, we have to define an $\mathcal{R}$-matrix acting between different Fock spaces $\mathcal{F}^{\alpha}_{x_1}\otimes \mathcal{F}^{\beta}_{x_2}$. In the following we will use the results of \cite{2015arXiv151208779B,Litvinov:2016mgi} and also \cite{feigin2020deformations} where various integrable systems of this type considered in details for the $q$-deformed case. To the Fock module $\mathcal{F}^{\alpha}_v$ we assign a free bosonic field \eqref{free-field}:
\begin{equation}
\partial\varphi(x)=-i\frac{v}{\sqrt{\epsilon_{\beta}\epsilon_{\gamma}}}+\sum\limits_{n\ne 0}a_n e^{-i n x},\qquad
 [a_{m},a_{n}]=m\delta_{m,-n},
\end{equation}
here $(\alpha,\beta,\gamma)=\textrm{perm}(1,2,3)$. To the tensor product of two Fock modules we have to assign a $W$-algebra and an $R$-matrix. If the both Fock modules are of the same type $\mathcal{F}^{\alpha}_{x_1}\otimes \mathcal{F}^{\alpha}_{x_2}$ then we assign to them two Screening currents:
\begin{equation}
    S^{\pm}_{\alpha}=\oint e^{ \left(\frac{\epsilon_{\beta}}{\epsilon_{\gamma}}\right)^{\pm\frac{ 1}{2}} (\varphi_1(x)-\varphi_2(x))}dx,
\end{equation}
where $(\alpha,\beta,\gamma)=\text{perm}(1,2,3)$. The $W$-algebra which commutes with these Screenings consists of two currents of spin $1$ and $2$
\begin{gather}
    W^{\alpha}_1=\partial \varphi_1(x)+\partial \varphi_2(x)\\
    W^{\alpha}_2=\frac{1}{2}(\partial\varphi_1(x)-\partial\varphi_2(x))^2+\frac{\epsilon_{\alpha}}{\sqrt{\epsilon_{\beta}\epsilon_{\gamma}}} (\partial^2\varphi_1(x)-\partial^2\varphi_2(x))
\end{gather}
defines the $R$-matrix in the usual way:
\begin{equation}
    \mathcal{R}_{1,2}(\varphi_1-\varphi_2|\epsilon_{\alpha},\epsilon_{\beta},\epsilon_{\gamma})W^{\alpha}_{1,2}=W^{\alpha}_{1,2}\Big|_{\varphi_1 \leftrightarrow \varphi_2} \mathcal{R}_{1,2}(\varphi_1-\varphi_2|\epsilon_{\alpha},\epsilon_{\beta},\epsilon_{\gamma}).
\end{equation}
Here $\mathcal{R}^{\alpha,\alpha}_{1,2}(\varphi_1-\varphi_2|\epsilon_{\beta},\epsilon_{\gamma},\epsilon_{\alpha})$ is our old Maulik-Okounkov  $\mathcal{R}$-matrix \eqref{RLL-Miura}.

Now to Fock modules of different types: $\mathcal{F}^{\alpha}_{x_1}\otimes \mathcal{F}^{\beta}_{x_2}$ we assign a single "fermionic" screening charge
\begin{equation}
    S_{f,{\gamma}}=\oint e^{\sqrt{\frac{\epsilon_{\alpha}}{\epsilon_{\gamma}}}\varphi_1(x)-\sqrt{\frac{\epsilon_{\beta}}{\epsilon_{\gamma}}} \varphi_2(x)} dx.
\end{equation}
This screening called "fermionic" because it is a zero mode of a free fermion: 
\begin{gather}
\psi(x)=e^{\sqrt{\frac{\epsilon_{\alpha}}{\epsilon_{\gamma}}}\varphi_1(x)-\sqrt{\frac{\epsilon_{\beta}}{\epsilon_{\gamma}}} \varphi_2(x)} \, , \quad \psi^{\dagger}(x)=e^{-\sqrt{\frac{\epsilon_{\alpha}}{\epsilon_{\gamma}}}\varphi_1(x)+\sqrt{\frac{\epsilon_{\beta}}{\epsilon_{\gamma}}} \varphi_2(x)}\\
\psi(x)\psi^{\dagger}(y)=\frac{1}{\sin(x-y)}+reg \\
S_{f,{\gamma}}=\psi_0
\end{gather} 
Corresponding $W-$algebra  which commutes with screening charges consists of two currents of spin $1$, $2$:
\begin{gather}
    W_{f;1}=\frac{1}{\sqrt{\epsilon_{\alpha}}}\partial\varphi_1(x)+\frac{1}{\sqrt{\epsilon_{\beta}}}\partial\varphi_2(x) \\
    W_{f;2}=(\partial \Phi(x))^2+\partial^2 \Phi(x),
\end{gather}
where $\Phi(x)=\sqrt{\frac{\epsilon_{\alpha}}{\epsilon_{\gamma}}}\varphi_1(x)-\sqrt{\frac{\epsilon_{\beta}}{\epsilon_{\gamma}}} \varphi_2(x)$ and also an auxiliary current of spin $3$. Again the $R$-matrix can be found from the condition:
\begin{equation}
\mathcal{R}_{1,2}^{\alpha,\beta} \ W_{f; 1,2}=P_{1,2}(W_{f; 1,2})\, \mathcal{R}_{1,2}^{\alpha,\beta},
\end{equation}
here $P_{1,2}$ is a permutation operator $P_{1,2}: \mathcal{F}^{\alpha}_{x_1}\otimes \mathcal{F}^{\beta}_{x_2}\to \mathcal{F}^{\beta}_{x_2}\otimes \mathcal{F}^{\alpha}_{x_1}$. Note that now we have to permute not only the bosonic field $\varphi_1 \leftrightarrow \varphi_2$, but also have to exchange $\epsilon_{\alpha}\leftrightarrow \epsilon_{\beta}$ (we use $(\alpha,\beta,\gamma)=\text{perm}(1,2,3)$).

Fermionic $R$-matrix has very similar form in terms of free fermions:
\begin{equation}
\mathcal{R}_{1,2}^{\alpha,\beta}=\exp\Big[\frac{1}{2\pi}\int_{0}^{2\pi}:\psi^{\dagger}(x)\log\left( \partial \right)\psi(x): dx\Big].
\end{equation}
\paragraph{Boundaries and $K$-matrices.}
We already seen \eqref{K1}-\eqref{K3} that there is three types of boundaries, which produce three types of $\mathcal{K}$-matrices. First let us consider the case of the right boundary. We will use the following notation $\mathcal{F}_{1}^{\alpha_1}\otimes\mathcal{F}_{2}^{\alpha_2}\dots\otimes\mathcal{F}_{n}^{\alpha_n}\Big|\beta_R$  for $n$ Fock spaces and the right boundary. The case of a left boundary is completely similar and can be obtained by the following isomorphism
\begin{equation}
(\beta_L=\beta_R)\Big|\mathcal{F}_{n}^{\alpha_n}\otimes\mathcal{F}_{n-1}^{\alpha_{n-1}}\dots\otimes\mathcal{F}_{1}^{\alpha_1} \simeq D_1 \dots D_n\left( \mathcal{F}_{1}^{\alpha_1}\otimes\mathcal{F}_{2}^{\alpha_2}\dots\otimes\mathcal{F}_{n}^{\alpha_n}\Big|\beta_R\right),    
\end{equation}
where $D_i$ is the operator of reflection of the bosonic fields $\varphi_i\to-\varphi_i$.

For the Fock module of type $\alpha$ and the boundary of type $\beta$:  $\mathcal{F}^{\alpha}_n\Big|\beta$ we assign two screenings charges:
\begin{equation}
    S_{\gamma}^{\pm}=\oint e^{\left(\frac{2\epsilon_{\beta}}{\epsilon_{\gamma}}\right)^{\pm \frac{1}{2}}\sqrt{2}\varphi_n(x)} dx,
\end{equation}
where $(\alpha,\beta,\gamma)=\text{perm}(1,2,3)$. The corresponding $K$-matrix is equal to:
\begin{equation}
\mathcal{K}_{\alpha|\beta}=\mathcal{R}(\sqrt{2}\varphi_n|\frac{\epsilon_{\beta}}{\sqrt{2}},\sqrt{2} \epsilon_{\gamma},-\frac{\epsilon_{\beta}}{\sqrt{2}}-\sqrt{2} \epsilon_{\gamma})
\end{equation}
If the Fock module and the boundary are of the same color, then the $K$-matrix is equal to the identity matrix:
\begin{equation}
    \mathcal{K}_{\alpha|\alpha}=\textrm{Id}.
\end{equation}

Additional screening charges depend not only on the last Fock module $\mathcal{F}^{\alpha}_{x_n}$, but on the previous one $\mathcal{F}^{\beta}_{x_{n-1}}$. For $\mathcal{F}_{n-1}^{\alpha}\otimes \mathcal{F}_n^{\alpha}\Big|\alpha$ we have
\begin{equation}
S^{\pm}_{\alpha}=\oint e^{ \left(\frac{\epsilon_{\beta}}{\epsilon_{\gamma}}\right)^{\pm\frac{ 1}{2}} (\varphi_{n-1}(x)-\varphi_n(x))}dx \,,\quad  \bar{S}^{\pm}_{\alpha}=\oint e^{ \left(\frac{\epsilon_{\beta}}{\epsilon_{\gamma}}\right)^{\pm\frac{ 1}{2}} (\varphi_{n-1}(x)+\varphi_n(x))}dx.
\end{equation}
And for $\mathcal{F}_{n-1}^{\alpha}\otimes \mathcal{F}_n^{\beta}\Big|\beta$
we have:
\begin{equation}
     S_{f,{\gamma}}=\oint e^{\sqrt{\frac{\epsilon_{\alpha}}{\epsilon_{\gamma}}}\varphi_{n-1}(x)-\sqrt{\frac{\epsilon_{\beta}}{\epsilon_{\gamma}}} \varphi_n(x)} dx \,, \quad \bar{S}_{f,{\gamma}}=\oint e^{\sqrt{\frac{\epsilon_{\alpha}}{\epsilon_{\gamma}}}\varphi_{n-1}(x)+\sqrt{\frac{\epsilon_{\beta}}{\epsilon_{\gamma}}} \varphi_n(x)} dx.
\end{equation}
Equivalently corresponding $W$-algebra may be found from the condition of symmetry under reflection of the last boson $\varphi_n\to -\varphi_n$. These rules may be summarised by the following picture:
\begin{equation*}
 \psfrag{F1}{\resizebox{0.78cm}{!}{$\mathcal{F}_{n-1}^{\alpha}$}}
 \psfrag{F2}{\resizebox{0.5cm}{!}{$\mathcal{F}^{\beta}_{n}$}}
 \psfrag{F3}{\resizebox{0.78cm}{!}{$\mathcal{F}^{\alpha}_{n-1}$}}
 \psfrag{F4}{\resizebox{0.5cm}{!}{$\mathcal{F}^{\alpha}_{n}$}}
 \psfrag{F5}{\resizebox{0.78cm}{!}{$\mathcal{F}^{\alpha}_{n-1}$}}
 \psfrag{F6}{\resizebox{0.78cm}{!}{$\mathcal{F}^{\alpha}_{n-1}$}}
 \psfrag{F7}{\resizebox{0.5cm}{!}{$\mathcal{F}^{\beta}_{n}$}}
 \psfrag{F8}{\resizebox{0.78cm}{!}{$\mathcal{F}^{\alpha}_{n-1}$}}
 \psfrag{F9}{\resizebox{0.5cm}{!}{$\mathcal{F}^{\alpha}_{n}$}}
 \psfrag{10}{\resizebox{0.3cm}{!}{$\alpha$}}
 \psfrag{e12}{\resizebox{3cm}{!}{${\sqrt{\frac{\epsilon_{\alpha}}{\epsilon_{\gamma}}}\varphi_{n-1}(x)-\sqrt{\frac{\epsilon_{\beta}}{\epsilon_{\gamma}}} \varphi_{n}(x)}$}}
 \psfrag{e56}{\resizebox{2cm}{!}{$\left(2\frac{\epsilon_{\beta}}{\epsilon_{\gamma}}\right)^{\pm\frac{1}{2}}\sqrt{2}\varphi_{n-1}(x)$}}
 \psfrag{e34}{\resizebox{3cm}{!}{$\left(\frac{\epsilon_{\beta}}{\epsilon_{\gamma}}\right)^{\pm\frac{1}{2}} (\varphi_{n-1}(x)-\varphi_{n}(x))$}}  
 \psfrag{e67}{\resizebox{3cm}{!}{${\sqrt{\frac{\epsilon_{\alpha}}{\epsilon_{\gamma}}}\varphi_{n-1}(x)-\sqrt{\frac{\epsilon_{\beta}}{\epsilon_{\gamma}}} \varphi_{n}(x)}$}}
 \psfrag{eb67}{\resizebox{3cm}{!}{${\sqrt{\frac{\epsilon_{\alpha}}{\epsilon_{\gamma}}}\varphi_{n-1}(x)+\sqrt{\frac{\epsilon_{\beta}}{\epsilon_{\gamma}}} \varphi_{n}(x)}$}}
  \psfrag{6}{\resizebox{0.3cm}{!}{$\beta$}}
    \psfrag{7}{\resizebox{0.3cm}{!}{$\beta$}}
    \psfrag{e89}{\resizebox{3cm}{!}{$\left(\frac{\epsilon_{\beta}}{\epsilon_{\gamma}}\right)^{\pm\frac{1}{2}} (\varphi_{n-1}(x)-\varphi_{n}(x))$}}  
     \psfrag{eb89}{\resizebox{3cm}{!}{$\left(\frac{\epsilon_{\beta}}{\epsilon_{\gamma}}\right)^{\pm\frac{1}{2}} (\varphi_{n-1}(x)+\varphi_{n}(x))$}} 
   \includegraphics[width=0.65 \textwidth]{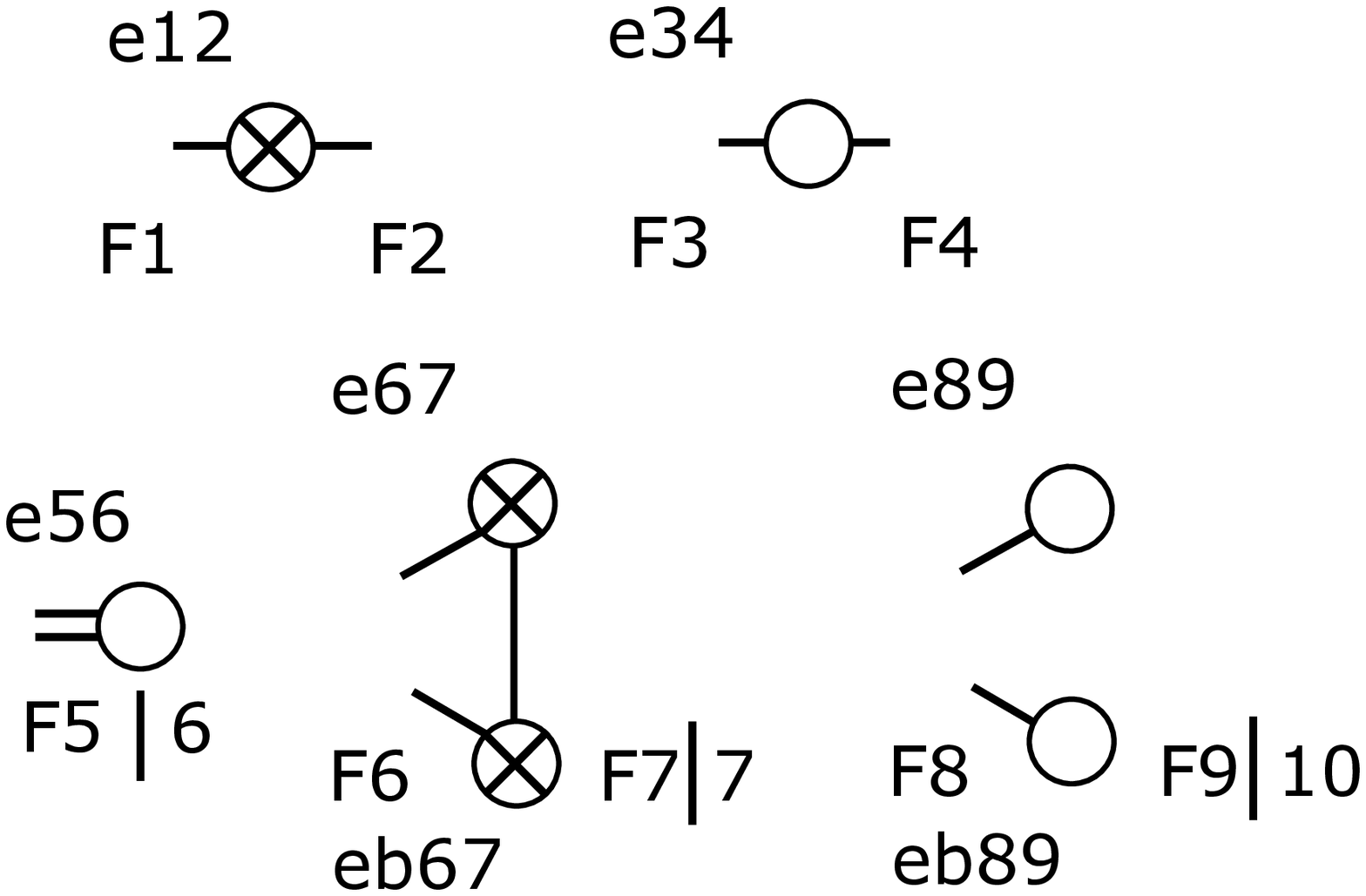}
\end{equation*}

Finally we may assign an integrable system to the chain of colored Fock spaces with two colored boundaries $\beta_L\Big|\mathcal{F}_1^{\alpha_1}\otimes \mathcal{F}_2^{\alpha_2} \dots \otimes \mathcal{F}_n^{\alpha_n}\Big|\beta_R$ , using the corresponding $R-$ and $K$-matrices. We may construct \textrm{KZ} Integrals of Motion \eqref{KZ_oper} and off-shell Bethe vectors \eqref{definition-off-shell-bethe-vector} in precisely the same way s described in the main text. We may also find local Integrals of Motion as commutant of screenings charges. The corresponding Bethe equations read as:
\begin{gather}
   r^{\beta_L}(x_i)r^{\beta_R}(x_i)A(x_i)A^{-1}(-x_i)\prod_{j\neq i}G(x_i-x_j)G^{-1}(-x_i-x_j)=1,
\\
G(x)=\frac{(x-\epsilon_1)(x-\epsilon_2)(x-\epsilon_3)}{(x+\epsilon_1)(x+\epsilon_2)(x+\epsilon_3)},\qquad
  A(x)=\prod_{k=1}^n\frac{x-u_k+\frac{\epsilon_{\alpha_k}}{2}}{x-u_k-\frac{\epsilon_{\alpha_k}}{2}},\quad r^{\alpha}(x)=-\frac{x+\epsilon_{\alpha}/2}{x-\epsilon_{\alpha}/2}.
\end{gather}


\bibliographystyle{MyStyle}
\bibliography{MyBib}

\end{document}